\documentclass[openacc]{rstransa}
\usepackage{bm}
\usepackage{caption}
\usepackage{subcaption}
\newcommand{\anna}[1]{{#1}}

\graphicspath{ {./images/} }

\begin{document}
\title{Transition to chaos and modal structure of magnetized Taylor--Couette flow}
\author{A.~Guseva and S.~M.~Tobias}
\address{Department of Applied Mathematics, University of Leeds}
\subject{...}
\corres{Anna Guseva\\
\email{A.Guseva@leeds.ac.uk}}

\begin{abstract}
Taylor-Couette flow is often used as a simplified model for complex rotating flows in the interior of stars and accretion disks. The flow dynamics in these objects is influenced by magnetic fields. For example, quasi-Keplerian flows in Taylor-Couette geometry become unstable to a travelling or standing wave in an external magnetic field if the fluid is conducting; there is an instability even when the flow is hydrodynamically stable. This magnetorotational instability leads to the development of chaotic states and, eventually, turbulence, when the cylinder rotation is sufficiently fast. The transition to turbulence in this flow can be complex, with the coexistence of parameter regions with spatio-temporal chaos and regions with quasi-periodic behaviour, involving one or two additional modulating frequencies. Although the unstable modes of a periodic flow can be identified with Floquet analysis, here we adopt a more flexible equation-free data-driven approach. We analyse the data from the transition to chaos in the magnetized Taylor-Couette flow and identify the flow structures related to the modulating frequencies with Dynamic Mode Decomposition; this method is based on approximating nonlinear dynamics with a linear infinite-dimensional Koopman operator. With the use of these structures, one can construct a nonlinear reduced model for the transition.
\end{abstract}

\begin{fmtext}
\section{Introduction}
Stability and transition to turbulence in fluid flows remains of interest for scientists since the beginning of the 20th century.  In 1923, Taylor explored a linearly unstable flow between two concentric rotating cylinders both theoretically and experimentally in his influential work \cite{taylor1923viii}.
Its starting points  was a combination of the Rayleigh stability criterion for inviscid rotating fluids,
\begin{equation}\label{eq:RayleighCriterion}
    \Omega_o R_o^2 < \Omega_i R_i^2 \quad \text{for instability,}
\end{equation}
\end{fmtext}
\maketitle

together with the viscosity measurements by Couette and Mallock, indicating an instability in an analogous setup. Here  angular velocities $\Omega_i$, $\Omega_o$ and radii $r_i$, $r_o$ correspond to the inner and outer cylinders. Taylor's  stability diagrams for axisymmetric disturbances are in an excellent agreement with  experiments though  the final pages of his work are devoted to the instability of the axisymmetric vortices themselves. He observed that ``a large increase [in the speed] caused the symmetric motion to break down into some kind of turbulent motion...", and that ``each vortex was pulsating so that its cross-section varied periodically".

Since then, Taylor--Couette flow (TCF) has also become a  model for turbulence generation in rapidly rotating astrophysical flows~\cite{rudiger2020large, guseva2017transport}, where turbulence is important for angular momentum transport, magnetic field generation and mixing of chemical species. In those studies, the velocity of the cylinders is set to approximate the desired astrophysical rotation, for example, Keplerian profile $\Omega \sim r^{-1.5}$ of an accretion disc.  Quasi-Keplerian flows, with their angular momentum increasing with radius, are hydrodynamically linearly stable to infinitesimal perturbations according to the Rayleigh criterion~\eqref{eq:RayleighCriterion}. Though transition to turbulence through finite perturbations at large rotation speeds can not be ruled out, Taylor--Couette flow is remarkably stable in quasi-Keplerian regimes, up to Reynolds numbers of $Re\sim 10^6$ in experiments \cite{ji2006hydrodynamic}. Thus, other physical  mechanisms of instability in quasi-Keplerian flows are frequently considered. One of them,  magnetorotational instability (MRI), arises in differentially rotating flows threaded by large-scale magnetic fields,  frequent in astrophysical objects.  Taylor--Couette flow was used as a model for experimental studies of MRI \cite{stefani2006experimental}.

The stability of Taylor-Couette flow in the presence of magnetic fields was first studied by Velikhov  \cite{velikhov1959stability} and Chandrasekhar \cite{chandrasekhar1981hydrodynamic} in the 1950s. Considering axisymmetric perturbations and axial magnetic field, Velikhov concluded that magnetized Taylor--Couette flow is unstable if 
\begin{equation}\label{eq:SMRICriterion}
    d \Omega^2/dr >0,
\end{equation}
i.e. when the angular velocity, and not angular momentum, decreases with radius.  In an ideally conductive fluid a radially displaced fluid parcel drags away the magnetic field line, ``glued" into the flow, and retains its previous angular velocity. In the new location, the fluid parcel experiences three forces: magnetic tension, the centrifugal force, and the equilibrium pressure gradient. If the velocity decreases outwards, the centrifugal force of the fluid element is larger than the pressure gradient, and in sufficiently weak fields this leads to instability. If the magnetic field is too strong, magnetic tension  stabilizes the flow. Similar  arguments can be invoked for  azimuthal magnetic field; here the flow stability depends on the radial shape of the field  \cite{velikhov1959stability,kirillov2013extending}. Hollerbach \textit{et al} \cite{hollerbach2010nonaxisymmetric} showed numerically that nonaxisymmetric disturbances with azimuthal wave number $m=1$  are the most unstable in this case.

Most of the existing MRI studies focused on asymptotic behaviour of instability and properties of fully developed turbulence. Transition from MRI to turbulence in Taylor--Couette flow was not investigated systematically. Guseva \textit{et al} \cite{guseva2015transition} found that MRI arises in a supercritical Hopf bifurcation, and then the flow undergoes a subcritical Hopf bifurcation to a chaos when the ratio of viscosity $\nu$  to magnetic diffusivity $\eta$ of the fluid is low, i.e. magnetic Prandtl number  $Pm = \nu/\eta \sim 10^{-6}$.  In plasma-like fluids with large $Pm \sim 1$, \cite{guseva2017azimuthal} reported a more complex scenario of transition, with the flow passing a succession of oscillatory  states as the strength of magnetic field varies. The present work aims to analyse how these oscillatory states appear and evolve using a data-driven approach. We employ the method of Dynamic Mode Decomposition (DMD) for identification of coherent structures in physical systems. DMD was developed by Schmid~\cite{schmid2010dynamic} as an alternative to costly iterative methods of global stability analysis, with direct applications to fluid flows. It can be interpreted as the generalization of global stability analysis for both numerical and experimental data, and results in a set of ``dynamic'' spatial modes and corresponding  eigenvalues. We will refer to them as DMD modes and DMD eigenvalues, respectively. For nonlinear systems, DMD  represents linear tangent approximation of the system's dominant dynamics. The theoretical significance of this linear approximation is closely related to the idea that the dynamics of a nonlinear system of finite dimensions can be represented by a linear infinite-dimensional Koopman operator~\cite{rowley2009spectral}. This operator propagates flow observables in time, and its eigenvalues and eigenvectors fully define the dynamics of the system; DMD can be viewed as the numerical approximation of this operator. Compared to other decomposition methods like  Principal Orthogonal Decomposition (POD), DMD is superior in identifying flow frequencies, and therefore is most  appropriate for the analysis of the mentioned above oscillatory states.

This paper is structured as following: first, we introduce our numerical setup for Taylor--Couette flow, and the qualitative description of the transition to turbulence. After that, we describe DMD method in more detail. We present the results of DMD analysis and identify the dynamical components related to the transition. Finally, we discuss the results and give an outlook on the possible future work.

\section{Description of the flow}
The equations describing the motion of an incompressible conducting fluid in the presence of magnetic fields are the Navier--Stokes and induction equation:
\begin{align}
\bm{u}_t + \bm{u} \cdot \nabla \bm{u}& = - \frac{1}{\rho} \nabla p + \frac{1}{\mu_0 \rho} (\nabla \times \bm{B}) \times \bm{B} + \nu \nabla^2 \bm{u} + \bm{f}, \label{eq:NSt}  \\ 
\bm{B}_t&= \nabla \times (\bm{u} \times \bm{B}) - \eta \nabla^2 \bm{B}, \label{eq:Ind} \\ 
 & \nabla\cdot \bm{u} = \nabla \cdot \bm{B} = 0. \label{eq:divfree}
\end{align}
Note the feedback of magnetic Lorentz force~$(\nabla \times \bm{B}) \times \bm{B}$ on the flow; this force is an essential component of the MRI. Thus, the flow is intrinsically nonlinear in velocity field $\bm{u}$ and magnetic field $\bm{B}$. 
 
The laminar solution to~\eqref{eq:NSt} in hydrodynamic Taylor--Couette flow is
\begin{equation}\label{eq:lamprof}
    V(r) = C_1 r + C_2/r, \quad C_1= \frac{\Omega_o r_o^2 - \Omega_i r_i^2}{r_o^2 - r_i^2}, \quad C_2 = \frac{(\Omega_i - \Omega_o) r_i^2 r_o^2}{ r_o^2 - r_i^2}.
\end{equation}
 The radius ratio of the cylinders was set to $r_i/r_o =0.5$, and the rotation rate to $\Omega_o/\Omega_i = 0.26$,  approximating a quasi-Keplerian profile with $\Omega \to r^{-2}$, but still fulfilling criterion~\eqref{eq:SMRICriterion}. The dimensionless parameters are Reynolds number  $Re = \Omega_i r_i d/\nu$ and Hartmann number $Ha = B_0 d/\sqrt{\sigma/\rho \nu}$ which compares the strength of the Lorentz force to the viscous force.  Setting $Pm=1$ implies that the dissipation of magnetic and velocity fluctuations takes place on the same scale, and leaves only two free parameters in the flow, $Re$ and $Ha$.
 
 We solve~\eqref{eq:NSt},~\eqref{eq:Ind} and~\eqref{eq:divfree} using direct numerical simulations (DNS) in Taylor--Couette geometry. The laminar velocity profile~\eqref{eq:lamprof}, and the azimuthal magnetic field $B_\phi =B_0 (1/r)$ are imposed as forcing terms. The code has spectral discretization in the axial and azimuthal directions $z$ and $\varphi$, and the radial coordinate $r$ is discretized using finite differences. The nonlinear terms are evaluated in the physical space  and are de-aliased using $3/2$ rule; More details of the numerical method can be found in \cite{guseva2015transition}. The axial wave numbers are set to $k = \alpha k'$, $k' = 0,1,...$, with $\alpha = 0.5$, equivalent to setting the length of the cylinder to $L_z = 2 \pi /\alpha = 4\pi$. The spatial resolution is $N_r, N_z, N_\phi = (40, 64, 16)$, where $N_z$ and $N_\varphi$ in the number of Fourier modes in respective directions. \anna{Each run is started from a small non-axisymmetric perturbation of  one  flow mode with $k' = 2$ ($\alpha k' = 1$), $m=1$. This initial condition simplifies  the flow dynamics, constraining it to the subspace with only even wave numbers, $k' = 0,2,4,...$.}  Without this constraint, the odd wave numbers also become active, which results in a considerably more complex transition scenario left for the future work. 

\subsection{Magnetorotational instability and transition to chaos in DNS}
\begin{figure}
     \centering
     \begin{subfigure}[b]{0.47\textwidth}
         \centering
         \includegraphics[width=\textwidth]{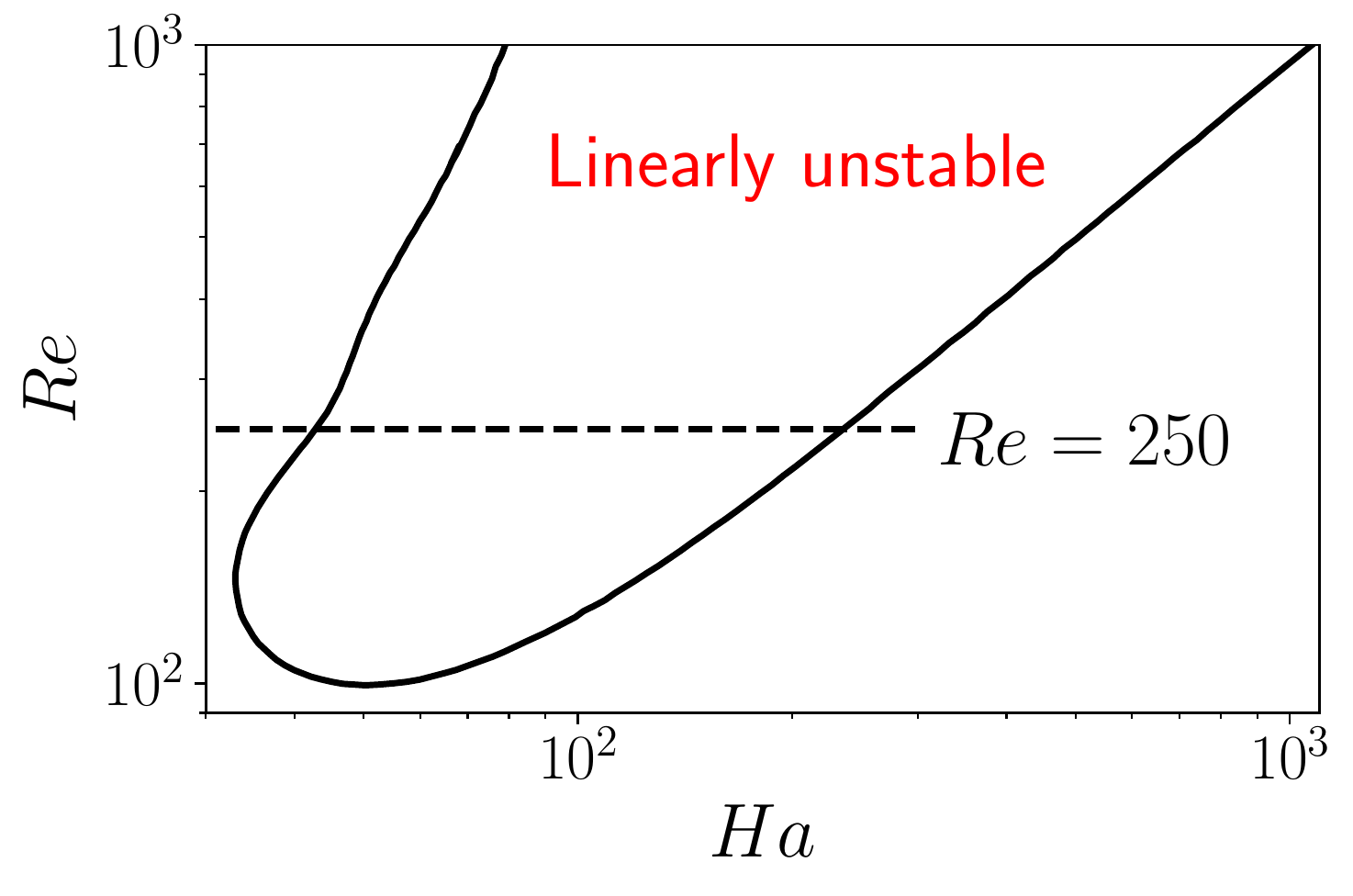}
         \caption{}
         \label{fig:linstab_map}
     \end{subfigure}
     \hfill
     \begin{subfigure}[b]{0.47\textwidth}
         \centering
         \includegraphics[width=\textwidth]{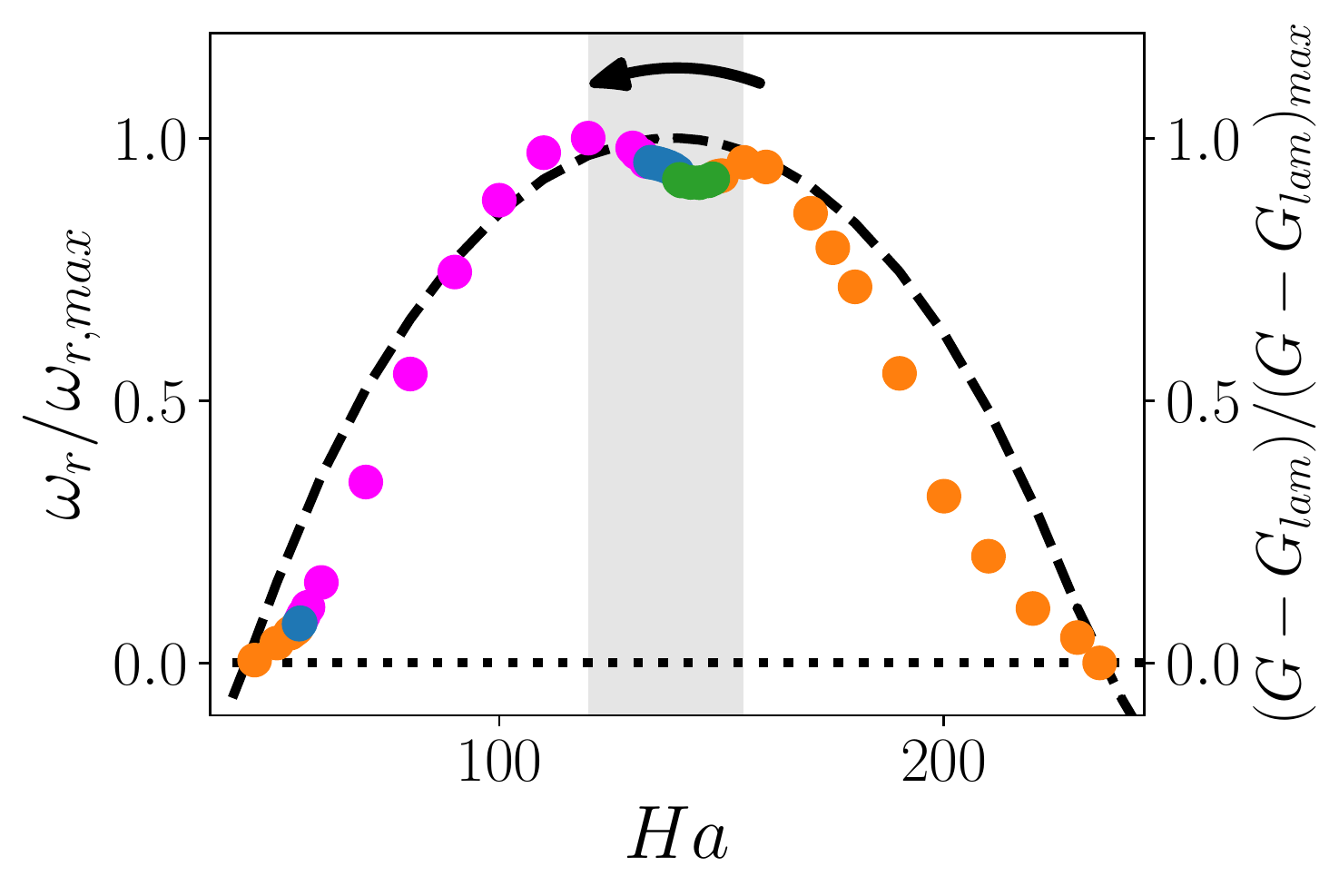}
         \caption{}
         \label{fig:varHa_Re250}
     \end{subfigure}
     \vfill
       \begin{subfigure}[b]{\textwidth}
        \centering
         \includegraphics[width=\textwidth]{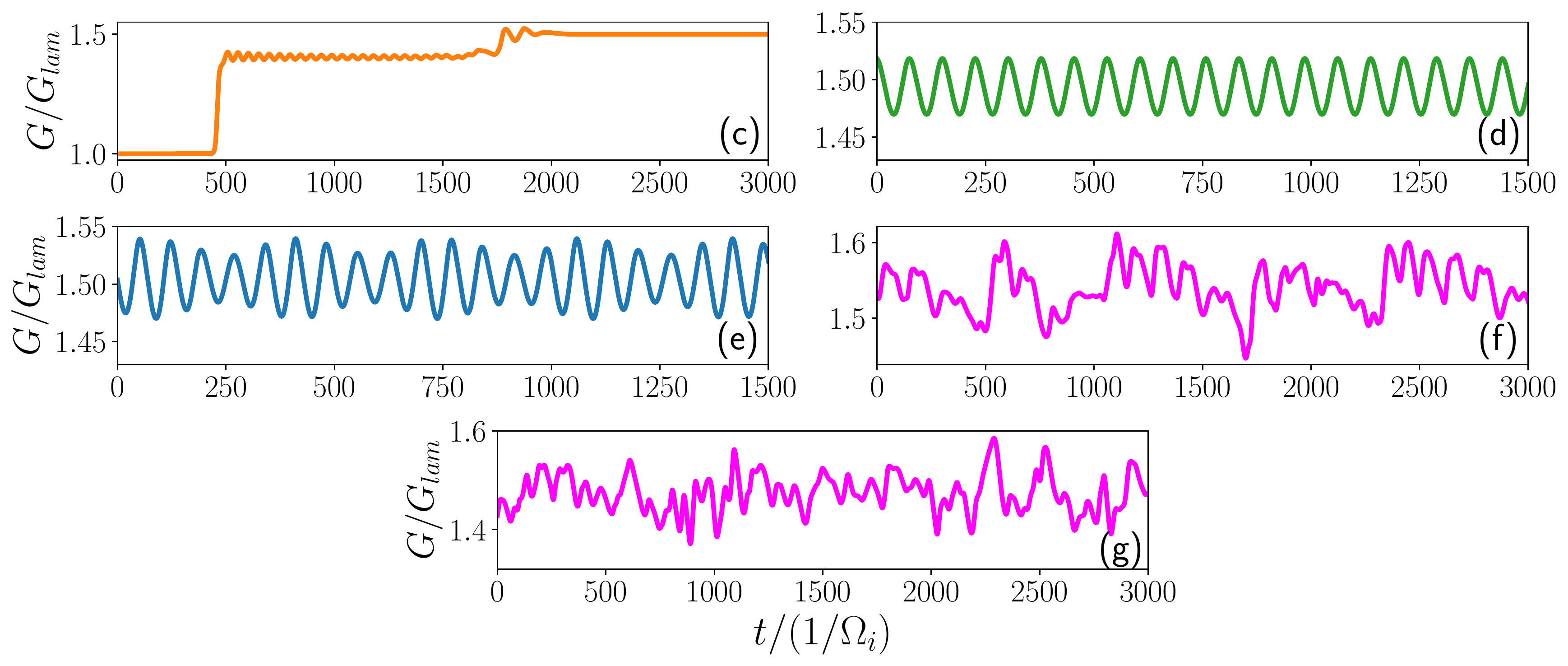}
        \label{fig:G_Ha149}
     \end{subfigure}
        \caption{(a) The linear stability map of the flow. (b) The growth rates of the instability, $\omega_r / \omega_{r, max}$, along the line $Re=250$, normalized with their maximum (dashed). Time-averaged torque with the laminar value subtracted, normalized with its maximum. Dotted line denotes $\omega_r =0$.  (c-g) Torque as a function of time. (c) $Ha=149$, standing wave ($Ha=50$ is analogous); (d) $Ha=145$, 1-frequency oscillation ; (e) $Ha=140$, 2-frequency oscillation;  (f) $Ha=120$, with chaotic but still relatively regular behaviour; (g) a chaotic flow at $Ha=100$. \anna{The line colours in panels (c-g) correspond to the color of the points in panel (b), denoting different flow states. \anna{The shaded region denotes the transition to chaos analysed in this work; the arrow helps to orient the narrative in section 2(b) and further.}}}
        \label{fig:flow_description}
\end{figure}

The linear stability analysis of the flow, performed by linearizing~\eqref{eq:NSt},~\eqref{eq:Ind} and with the numerical method from \cite{hollerbach2010nonaxisymmetric}, shows the parameters  $Re$, $Ha$ where MRI is active (figure~\ref{fig:linstab_map}). The instability arises when $Re>Re_{cr} \approx 100$; above this threshold, magnetic field should be neither weak nor too strong for the flow to be unstable. As the Reynolds number increases, the magnetic field strength required to trigger instability also increases; however, the instability range becomes wider overall.  We focus on the case with $Re=250$, as in \cite{guseva2017azimuthal}.  The instability growth-rates $\omega_r$ along this line are represented by the dashed line in figure~\ref{fig:varHa_Re250}.

In DNS, the diagnostic quantity for the onset of instability is the friction torque on the cylinders $G$. It is related to transverse current of azimuthal motion in radial direction $J^\omega$ \cite{eckhardt2007torque} through 
\begin{equation}\label{eq:GJw}
    G \sim \nu^{-2} J^\omega, \qquad J^\omega = r^2 \left[ \langle u_r u_\phi \rangle_{A} - r \partial_r \langle \frac{u_\phi}{r} \rangle_{A} - \frac{Ha^2}{Pm} \langle B_r B_\phi \rangle_{A} \right],
\end{equation}
 where the angular momentum can be transported through the tension of magnetic field lines via the  Maxwell stress component $ B_r B_\phi$ \cite{guseva2017transport}. Here $\langle ... \rangle_A$ denotes a spatial average along a cylindrical surface $A$. $J^\omega$ is constant across the radius $r$, so that $G_i = G_o =G$  on average  for statistically steady flows. 
 In the absence of MRI, the laminar flow torque $G_{lam}$ is constant in time and can be calculated analytically from~\eqref{eq:lamprof}. As the instability develops, the friction on the cylinders increases. This increase is directly related to the dissipation enhancement in the flow \cite{lewis1999velocity}, as more energy is required to maintain the rotation. Figure~\ref{fig:varHa_Re250} shows a  correlation between the increase in $G$ and the instability growth rate $\omega_r$. Both $G$ and $\omega_r$ reach their maximum at about $Ha=120$, however, $G$ is not monotonic, with a local minimum developing where $\omega_r$ is the largest. Finally, the instability ceases to exist at about $Ha=230$ as the magnetic tension becomes too strong and stability is restored. 
 
\begin{figure}
     \centering
        \begin{subfigure}[b]{0.13\textwidth}
         \centering
         \includegraphics[width=\textwidth]{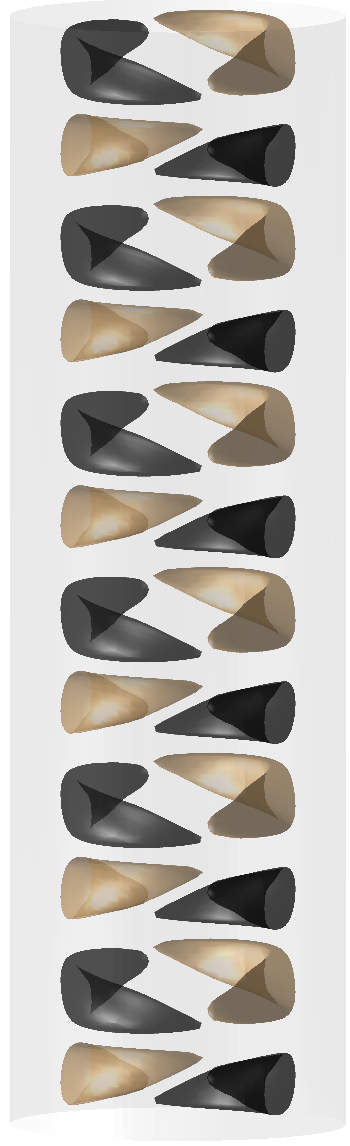}
         \caption{}
         \label{fig:vz_iso_Ha149}
     \end{subfigure}
    \hfill
         \begin{subfigure}[b]{0.13\textwidth}
         \centering
         \includegraphics[width=\textwidth]{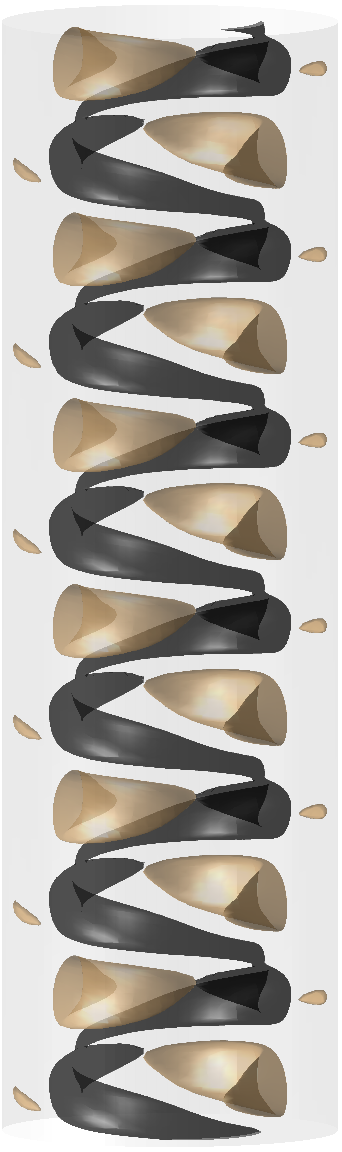}
         \caption{}
         \label{fig:vz_iso_Ha145}
     \end{subfigure}
     \hfill
         \begin{subfigure}[b]{0.13\textwidth}
         \centering
         \includegraphics[width=\textwidth]{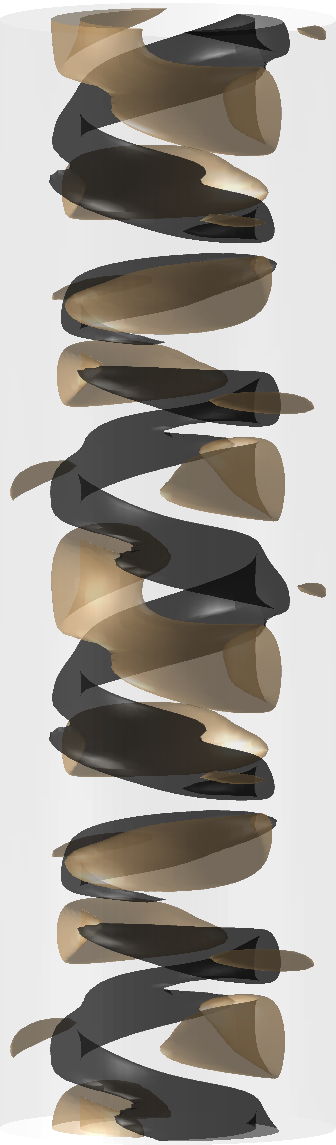}
         \caption{}
         \label{fig:vz_iso_Ha140}
     \end{subfigure}
         \hfill
         \begin{subfigure}[b]{0.13\textwidth}
         \centering
         \includegraphics[width=\textwidth]{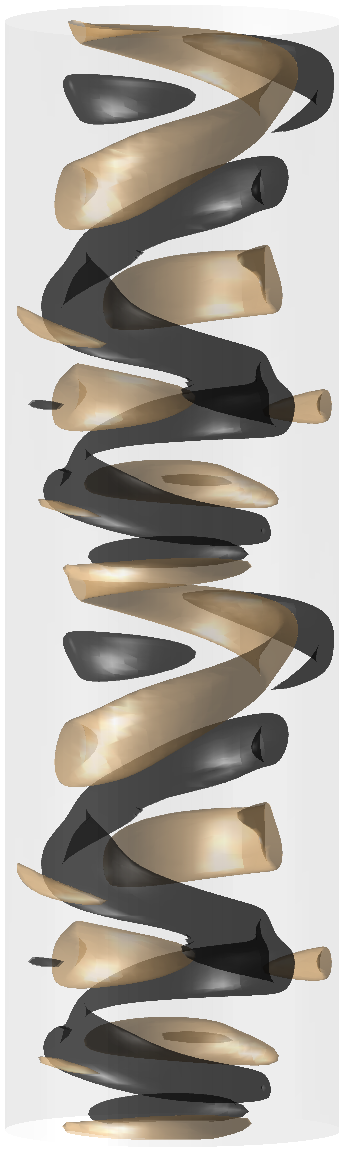}
         \caption{}
         \label{fig:vz_iso_Ha100}
     \end{subfigure}
          \hfill
         \begin{subfigure}[b]{0.13\textwidth}
         \centering
         \includegraphics[width=\textwidth]{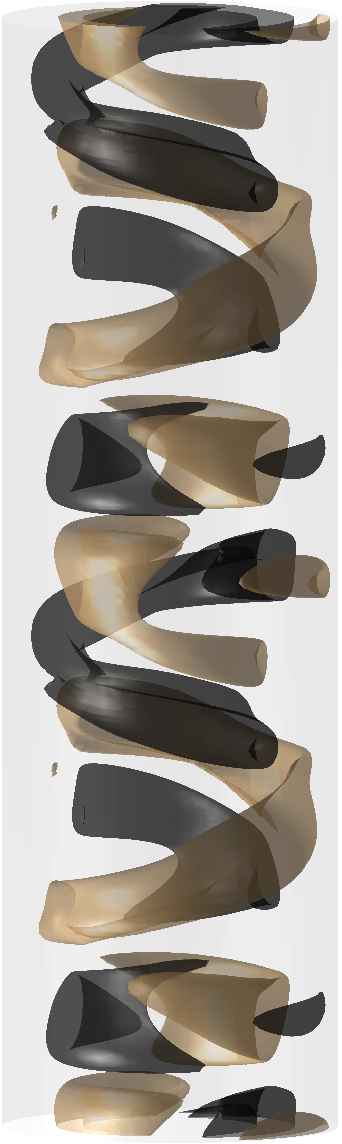}
         \caption{}
         \label{fig:bz_iso_Ha100_ch}
     \end{subfigure}
         \hfill
         \begin{subfigure}[b]{0.13\textwidth}
         \centering
         \includegraphics[width=\textwidth]{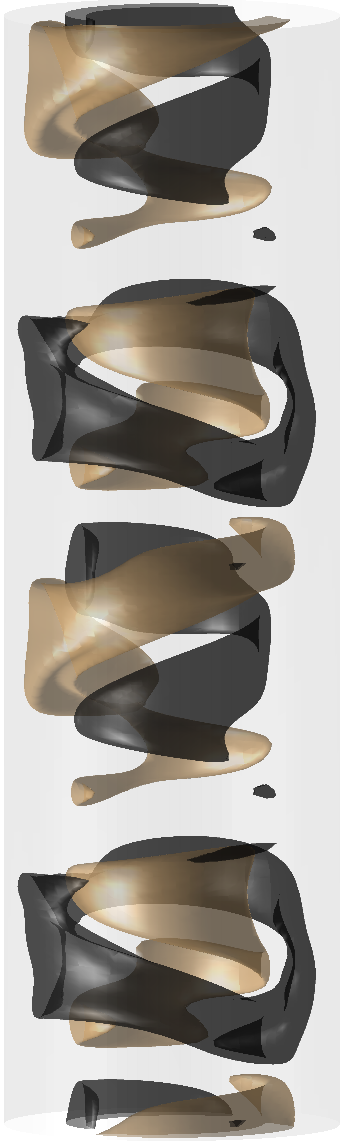}
         \caption{}
         \label{fig:bz_iso_Ha100_k4}
     \end{subfigure}
        \caption{Instantaneous snapshots of the vertical component of velocity field $v_z = \pm 12 [\nu /d]$. (a) $Ha=149$, standing wave  ($Ha=50$ is analogous), (b) $Ha=145$, 1-frequency flow state, (c) $Ha=140$, 2-frequency flow state, (d) $Ha=100$, chaotic solution. (e,f) Instantaneous snapshots of the vertical component of magnetic field $B_z$ at different times, $Ha=100$. The flow structures rotate azimuthally \anna{in the prograde direction}. }
        \label{fig:vz_snapshots}
\end{figure}

Now we focus our attention on the temporal behaviour of $G$.
 The instability appears first as a standing wave at $Ha\approx 50$ (figure~\ref{fig:vz_iso_Ha149}), which corresponds to a time-independent friction and energy state once initial transients saturate (figure~\ref{fig:flow_description}c). With increase in $Ha$, the flow rapidly becomes chaotic, with an abrupt transition to chaos at low $Ha$ and only a narrow interval of doubly-periodic in $G$ solutions. At $Ha=100$  the velocity and magnetic fields  exhibit chaotic features (figure~\ref{fig:flow_description}g), but retain some spatial structure (figures \ref{fig:vz_iso_Ha100},\ref{fig:bz_iso_Ha100_ch}). The vertical component of velocity and magnetic field is periodically dominated by a large-scale structure (figure~\ref{fig:bz_iso_Ha100_k4}). As $Ha$ is increased further, the chaotic behaviour begins to regularize, and  a slow modulation becomes discernible at $Ha=120$ (figure~\ref{fig:flow_description}f). Soon, the torque timeseries is nearly non-chaotic and \anna{again} shows doubly-periodic behaviour (figure~\ref{fig:flow_description}e), with a rapid oscillation of $\omega/\Omega_i \approx 0.09 $, and the slower modulation of $\omega/\Omega_i \approx 0.02$. Neither of these frequencies corresponds to the frequency of the MRI mode which rotates much faster azimuthally, at $\omega/ \Omega_i \in (0.3, 0.4) $. However, the velocity isosurfaces still show defects in figure~\ref{fig:vz_iso_Ha140}. The amplitude of the slower modulation of $G$ decreases with $Ha$, until only the rapid oscillation remains  at $Ha=141$ (figure~\ref{fig:flow_description}d). The spatial structure of the corresponding flow field is much more regular (figure~\ref{fig:vz_iso_Ha145}).  The magnitude of this oscillation also reduces with further increase in $Ha$, until it ceases to exist at $Ha=149$, and the steady state is again a standing wave (figure~\ref{fig:flow_description}c). \anna{ The strong asymmetry in transition to chaos (abrupt on the left, gradual on the right) is possibly related to subcriticality of the left stability border of the MRI \cite{guseva2017transport}, although further work is necessary to confirm this.} Table~\ref{tab:flow_regimes} gives an overview of the $Ha$ intervals of  the transition from chaos to regular behaviour.  In the following, we will focus on this parameter region, decreasing $Ha$ from about $Ha=150$ to $Ha = 100$, so that the flow complexity increases. 
 \begin{table}[t]
\centering
 \begin{tabular}{c| c |c |c|c} 
Range of $Ha$ & $(100, 133)$ &$(134, 140)$&$(140.5, 148)$ & $(149, 155)$\\
\hline
Behaviour of $G$, $E_{kin}$, $E_{mag}$ & Chaotic & 2-periodic & Periodic & Standing wave \\
 \end{tabular}
 \caption{Transition from MRI turbulence to a flow with only dominant unstable MRI mode, corresponding to the shaded region in figure~\ref{fig:varHa_Re250}. $E_{kin}$, $E_{mag}$ are the kinetic and magnetic energies of the flow, integrated over the computational domain.}\label{tab:flow_regimes}
\end{table}

\begin{figure}
     \centering
      \begin{subfigure}[b]{0.45\textwidth}
         \centering
         \includegraphics[width=\textwidth]{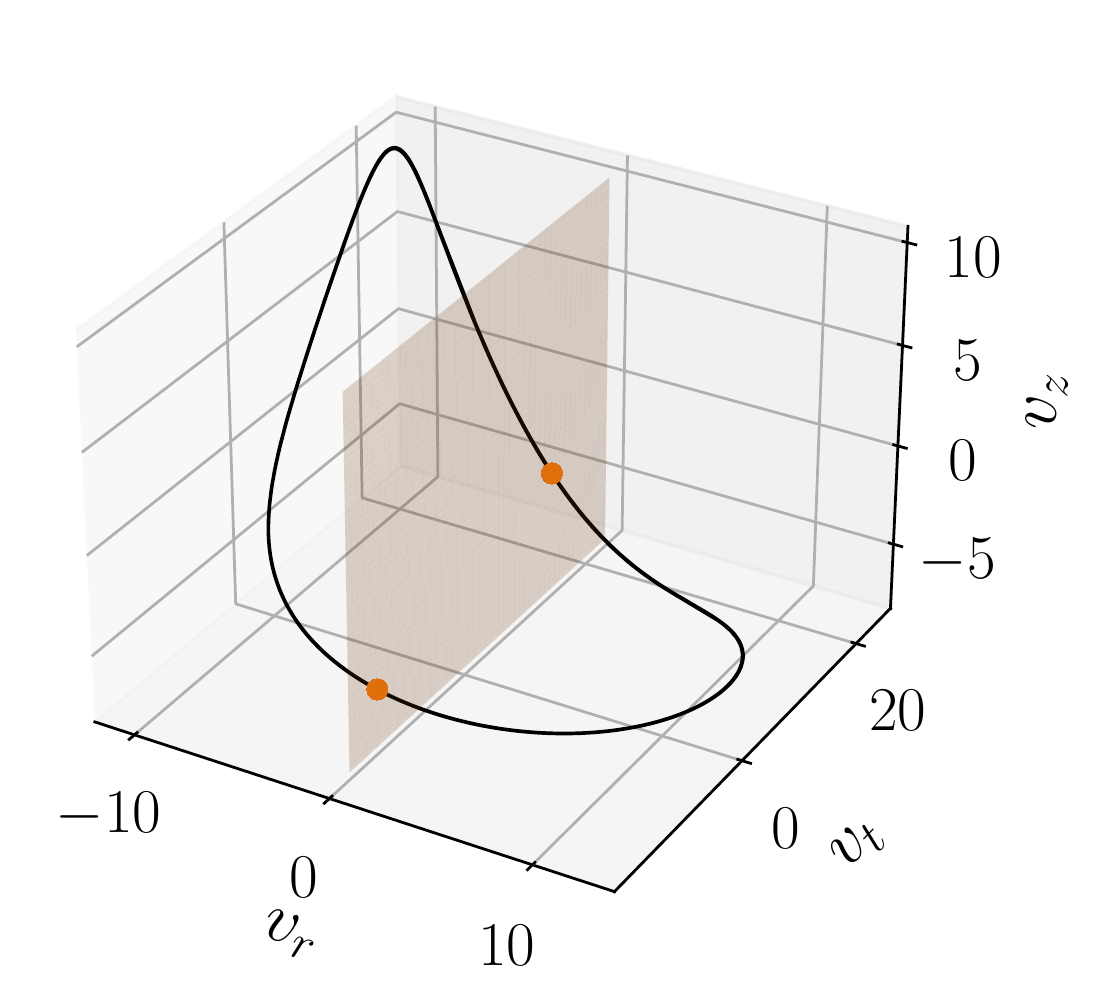}
         \caption{}
         \label{fig:vrtz_Ha149_5}
     \end{subfigure}
          \hfill
    \begin{subfigure}[b]{0.47\textwidth}
         \centering
         \includegraphics[width=\textwidth]{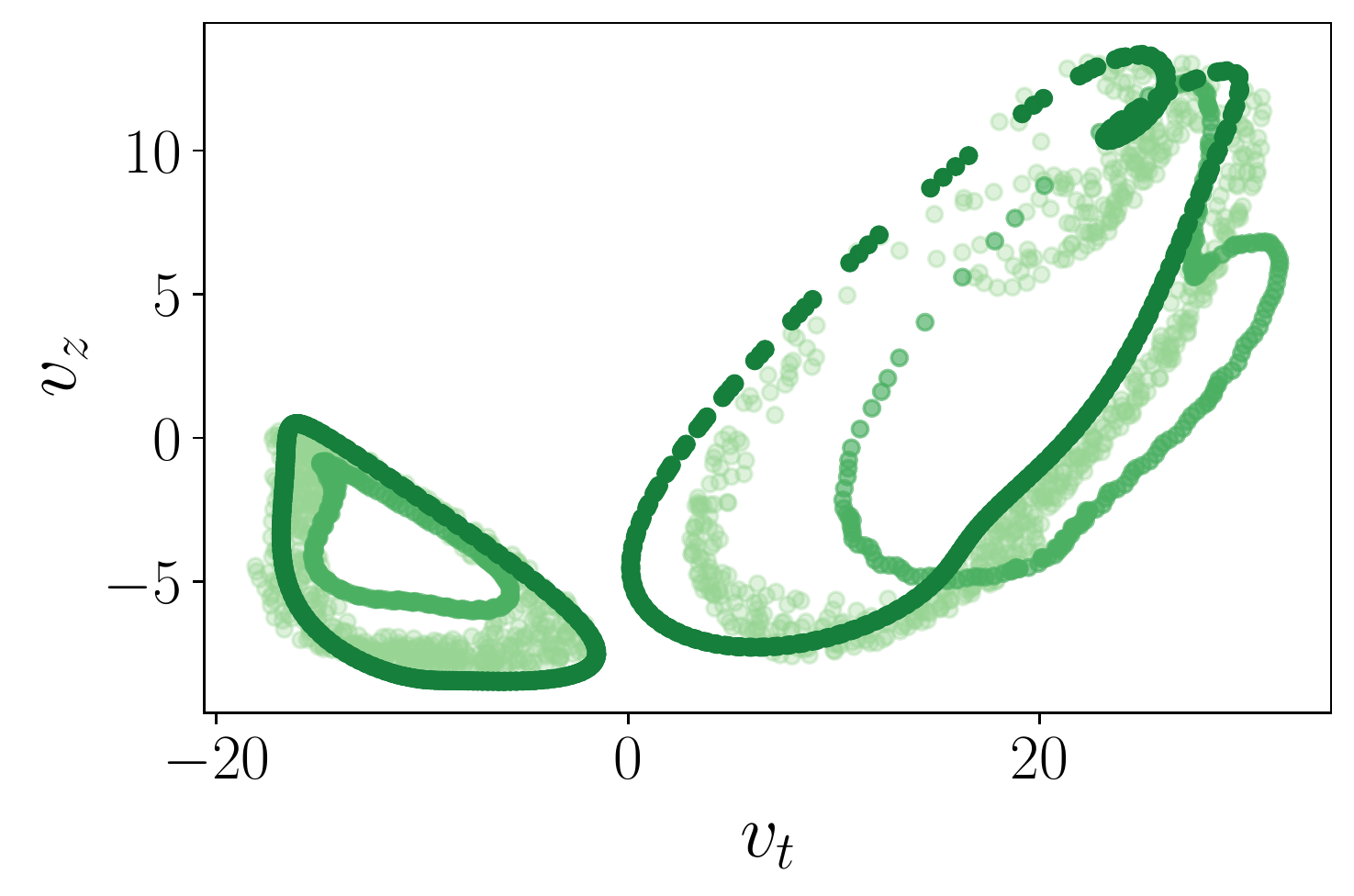}
         \caption{ }
         \label{fig:1tof2freq_poincare}
     \end{subfigure}
          \vfill
    \begin{subfigure}[b]{0.32\textwidth}
         \centering
         \includegraphics[width=\textwidth]{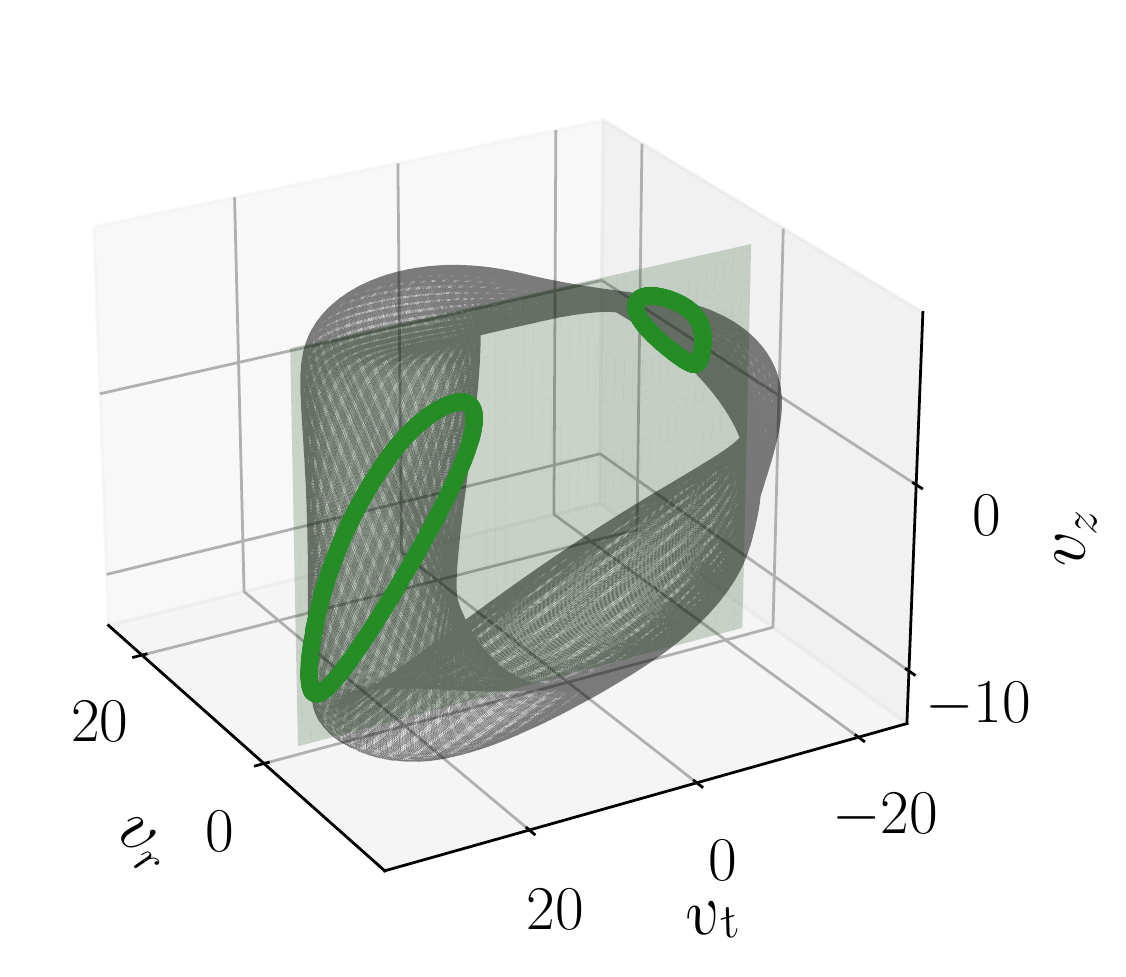}
         \caption{}
         \label{fig:vrtz_Ha147}
     \end{subfigure}
     \hfill
     \begin{subfigure}[b]{0.32\textwidth}
         \centering
         \includegraphics[width=\textwidth]{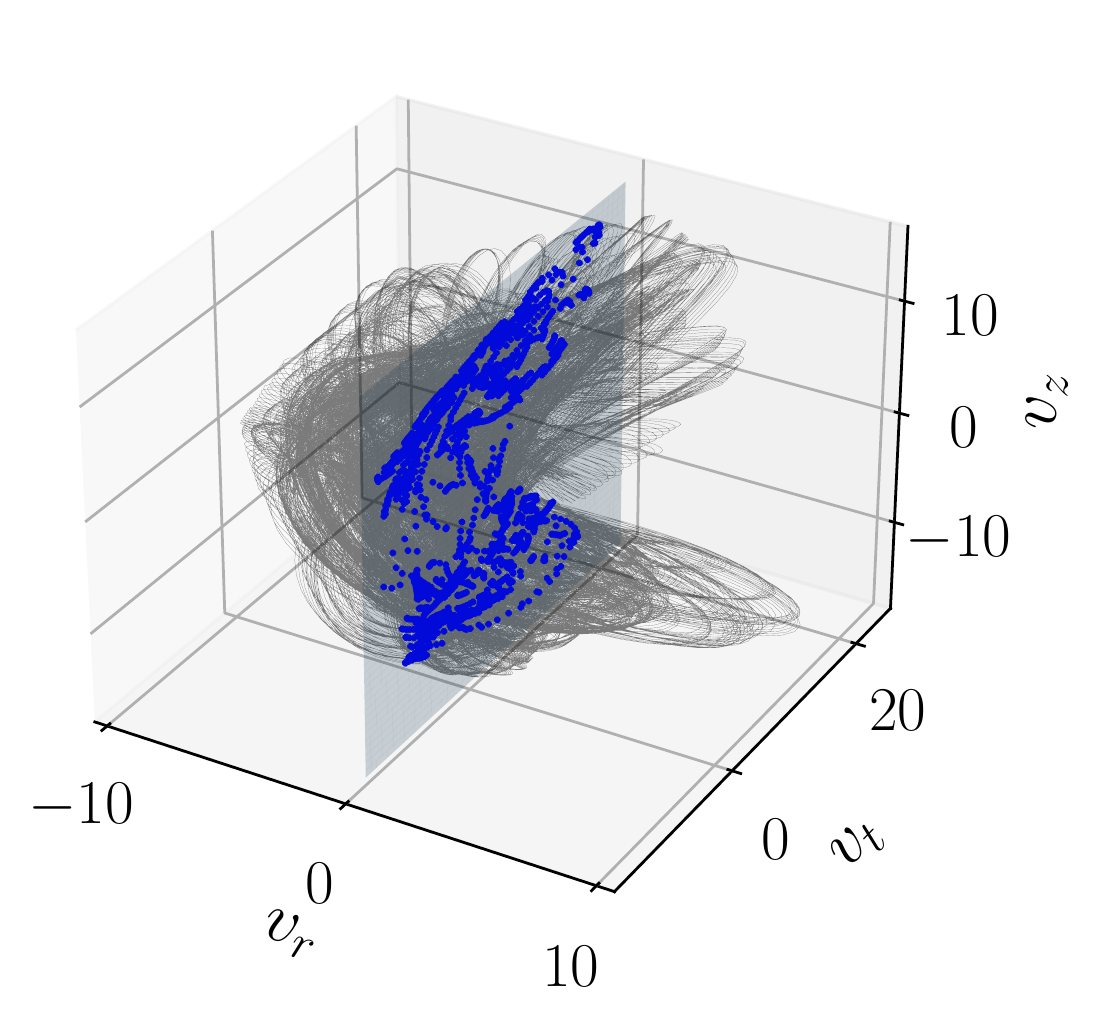}
         \caption{}
         \label{fig:vrtz_Ha140}
     \end{subfigure}
             \hfill
        \begin{subfigure}[b]{0.32\textwidth}
         \centering
         \includegraphics[width=\textwidth]{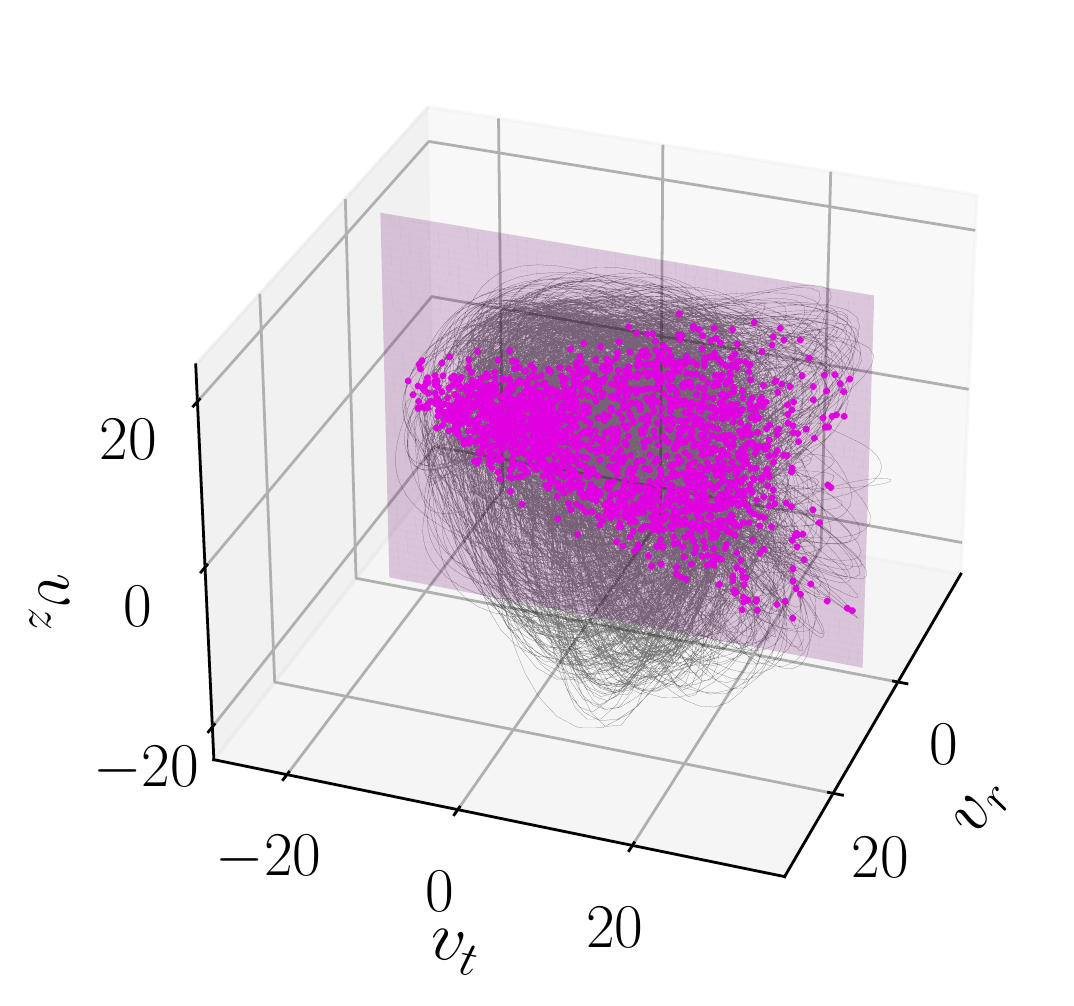}
         \caption{}
         \label{fig:vrtz_Ha100}
     \end{subfigure}
     \vfill 

    \begin{subfigure}[b]{0.32\textwidth}
         \centering
         \includegraphics[width=\textwidth]{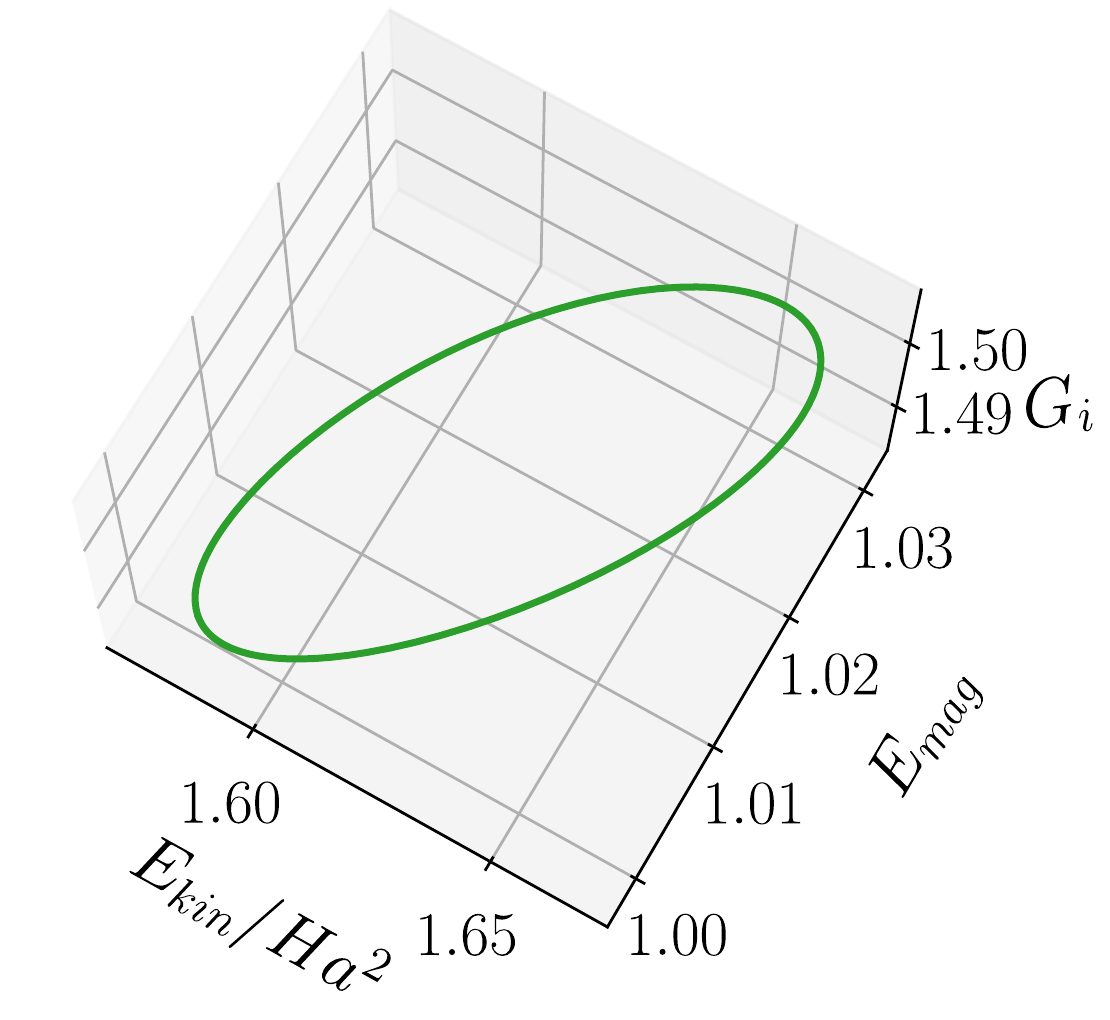}
         \caption{}
         \label{fig:GEkm_Ha147}
     \end{subfigure}
\hfill
          \begin{subfigure}[b]{0.32\textwidth}
         \centering
         \includegraphics[width=\textwidth]{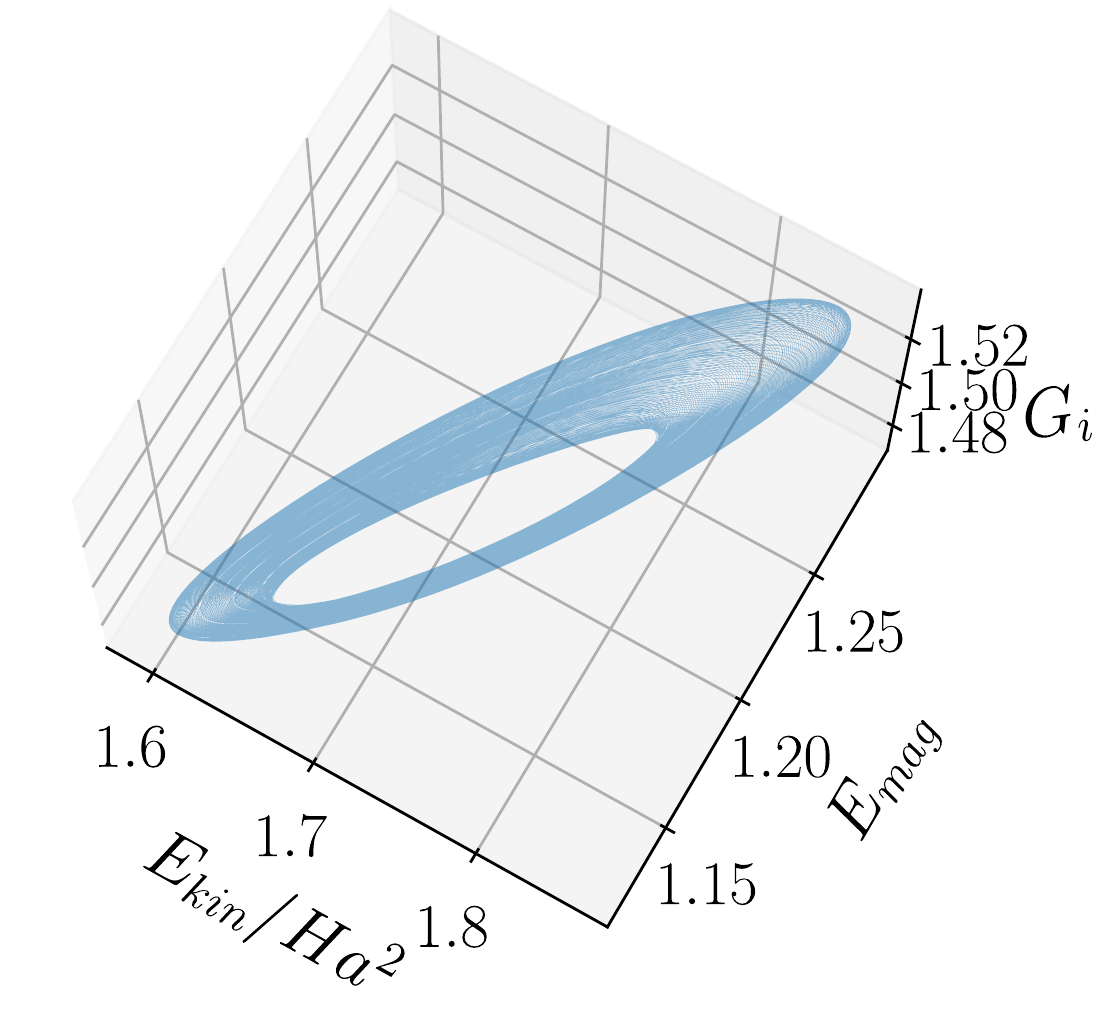}
         \caption{}
         \label{fig:GEkm_Ha140}
     \end{subfigure}
     \hfill
          \begin{subfigure}[b]{0.32\textwidth}
         \centering
         \includegraphics[width=\textwidth]{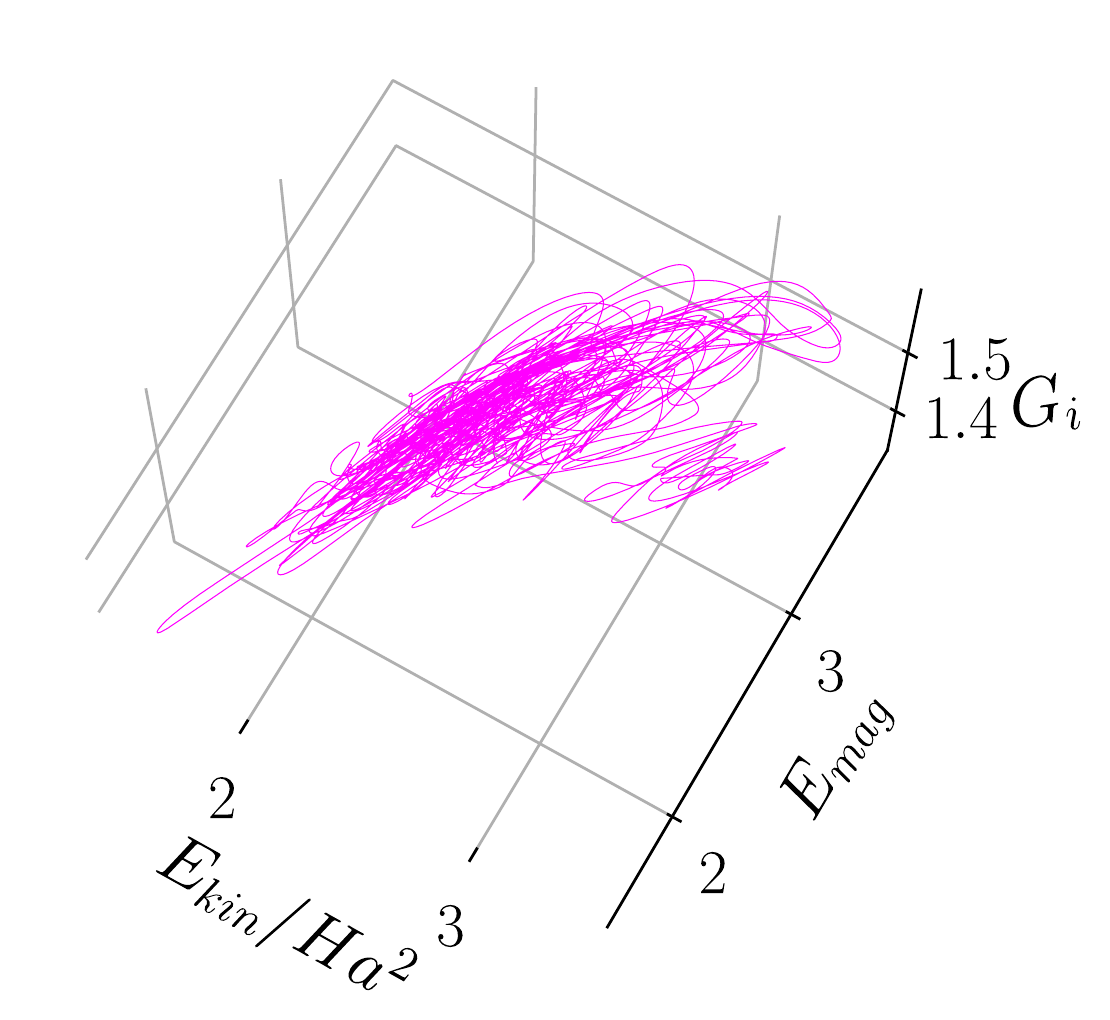}
         \caption{}
         \label{fig:GEkm_Ha100}
     \end{subfigure}
        \caption{Phase space maps and Poincare sections of the flow. (a) $Ha = 149$, standing wave.  (b) Transition between $1-$ and $2-$frequency solution through wrinkling of the chaotic attractor. Colours, from dark to light: $Ha\in [143,141, 140.5]$. 
        (c,f) $Ha=147$, periodic oscillation in $G$; (d,g) $Ha=140$, doubly-periodic $G$; (e,h) $Ha=100$, chaotic flow. \anna{$E_{kin}$ was normalized by $Ha^2/Pm$, with  $Pm=1$, to have the same units as $E_{mag}$.}}
        \label{fig:phase_space}
\end{figure}

\subsection{Chaos in phase space}
First, we explore the outlined above transition in phase; in particular the evolution of velocity $(u_r, u_\phi, u_z)$ at an arbitrary point in space $(r, \phi,z ) = (1.5,0,0)$. We subtract the mean from the three velocity components and draw a Poincare section through the plane $u_r = 0$ (figure~\ref{fig:phase_space}).  The second type of the phase space plot shows the time evolution of torque, kinetic and magnetic energy, $(G_i, E_{kin}, E_{mag})$, with $G_i$ a measure of dissipation.  The standing wave at $Ha=149$ in figure~\ref{fig:vz_iso_Ha149} rotates in azimuthal direction, so the velocities at a point oscillate periodically in figure~\ref{fig:vrtz_Ha149_5} --- this a removable frequency.  The standing wave is invariant in the $\phi$ and $z$-directions, so the torque and energies do not change in time and the integral flow state can be characterized as a fixed point (not shown). At about $Ha=148$,  $G_i$, $E_{kin}$ and $E_{mag}$ become periodic, and form  a periodic orbit in the phase space, as shown in figure~\ref{fig:GEkm_Ha147}. A second frequency appears in the system, and hence the corresponding system portrait in $u_{r,\varphi, z}$ is a two-torus  (figure~\ref{fig:vrtz_Ha147}), with velocity and magnetic field exhibiting a modulation. The intersection of the torus with Poincare section forms two closed loops, with increasing amplitude as we move away from the bifurcation point. When $Ha$ decreases further, the torus becomes more twisted \anna{and its sides visibly become closer compared to figure~\ref{fig:vrtz_Ha147}; figure~\ref{fig:1tof2freq_poincare} shows that} ``wrinkles" develop on its boundaries when $Ha \in (141,140)$. Finally, the torus breaks down, and the velocity and magnetic field lose their temporal coherency at $Ha=140$ (figure~\ref{fig:vrtz_Ha140}). The axial direction in the flow remains nevertheless less chaotic then the others, with an upward and a downward directions of motion forming in $u_z$ and $B_z$. The attractor in figure~\ref{fig:vrtz_Ha140} can be thought of as a sequence of twisted quasi-periodic orbits overlapping each other as they shift up or down in $u_z$.  The integral phase space features doubly-periodic oscillations and itself forms a relatively flat torus (figure~\ref{fig:GEkm_Ha140}). The quasi-periodic behaviour is maintained until $Ha=133$, when the flow becomes fully chaotic (figure~\ref{fig:GEkm_Ha100}). The orientation of the chaotic attractor is nevertheless preserved. The intersection of the flow trajectories with the Poincare section is denser, yet the central region of the attractor is less frequently revisited by the flow, and there still a reminiscence of the two lobes of high and low $u_z$ in figure~\ref{fig:vrtz_Ha100}.  Transient excursions away from the attractor occasionally occur, as visible by a trajectory excursion to a state with higher kinetic energy in figure~\ref{fig:GEkm_Ha100}. Overall, the flow transition to chaos through the breakdown of
a torus falls into the Ruelle--Takens scenario of transition to turbulence \cite{ruelle1971nature,manneville1995dissipative}. Nevertheless, here even chaotic flow states retain some regularity, and therefore could be potentially described with a few relevant dynamical components. In the next section, we will use the data-driven method of Dynamic Mode Decomposition to approach this problem.

\section{Dynamic mode decomposition (DMD)}
 Consider the system~\eqref{eq:NSt},\eqref{eq:Ind} in a general form,
\begin{equation}
\frac{d \bm{q}}{d t} = \bm{f}(\bm{q}, t, \bm{\mu}), \quad \bm{q}(t) = (\bm{u}, \bm{B}) , \quad \bm{\mu} = Re, Ha, Pm...
\end{equation}
where $\bm{f}$ is a nonlinear operator. We seek the best linear approximation to this nonlinear system in the form of
\begin{equation}\label{eq:DMDcont}
\frac{d \bm{q}}{d t} \approx \hat{A} \bm{q},  \quad \text{with solution} \quad \bm{q}(t) = \sum_{j=1}^n \phi_j \exp(\omega_j t) b_j.
\end{equation}
In general, the eigenvalues  of~\eqref{eq:DMDcont} are complex, i.e. $\omega = \omega_r + \mathrm{i} \omega_i$. In the simulations, the information about the flow is available in the form of magnetic and flow field snapshots $\bm{q}$, sampled every $\Delta t$ in time, so it is more practical to seek a discrete-time system
\begin{equation}\label{eq:DMDdisc}
\bm{q}_{k+1} = \exp(\hat{A} \Delta t) \bm{q}_k = A \bm{q}_k, \quad \text{with solution} \quad \bm{q}_k = \sum_{j=1}^n \phi_j \lambda_j^k b_j.
\end{equation}
The systems~\eqref{eq:DMDdisc} and~\eqref{eq:DMDcont} are analogous. Using definition~\eqref{eq:DMDdisc}, we implement the exact DMD algorithm \cite{tu2014dynamic}, as follows: 
\begin{enumerate}
\item Collect snapshots $\bm{q_k}$ of the system at timesteps $k = 1,2, \dots, K$
\item Construct data matrices $ Q =[q_1  \quad q_2 \cdots q_{m-1}]$,  $ Q' =[q_2 \quad q_2 \cdots q_m]$,  seeking $Q' \approx A Q$ 
 \item Compute the singular value decomposition (SVD): $Q = U \Sigma V^*$ \label{SVD}
  \item Keep $r$ modes from SVD  and define the reduced matrix $A_r = U_r^* Q' V_r \Sigma^{-1}_r$ \label{Ar}
  \item Solve eigenvalue problem $A_r \psi = \lambda \psi$
  \item Reconstruct DMD modes as $\varphi = \frac{1}{\lambda} Q' V_r \Sigma^{-1}_r \psi$ and frequencies as $\omega = \ln(\lambda)/ \Delta t$
\end{enumerate}
When approximation~\eqref{eq:DMDdisc} is valid, and the flow is steady-state, the dominant DMD modes are expected to be nearly neutral, with $\omega_r \to 0$, $|\lambda| \to 1$.
Step~\ref{Ar} introduces the  truncation parameter $r$, typically  defined by some criterion in the spectrum of singular values \anna{$\sigma$ in the diagonal of  matrix $\Sigma$ in~\ref{SVD}}. By increasing this parameter,  more dynamical information about the system may be kept. However, as singular values in decomposition~\ref{SVD} decreases, the singular vectors associated with these small singular values are increasingly linearly dependent and including them in~\ref{Ar} would make the subsequent decomposition ill-conditioned \cite{schmid2010dynamic}. A common choice is a $99\%$ cutoff of the SVD spectrum $\sigma$. In our case, this criterion leads to a very large intractable dynamical basis for the flow. Since the singular values spectrum $\sigma^2$ represent the energy content of the POD modes $U$, we re-define the cut-off parameter $r$ so that $ \sum_r \sigma^2_r / \sum \sigma^2 = 99 \%$, with the modes retaining 99\% of the energy of the respective quantity. Different components of magnetic and flow field have different cut-off values, depending on their spatial complexity. In general, $u_r$, $B_r$ have the largest modal basis, and $u_\varphi$, $B_\varphi$, influenced heavily by their mean fields, remain low-dimensional. For the latter, we set the cut-off parameter at $r=3$, including the mode with $\omega_i =0$, corresponding to non-oscillating motion, and two main complex conjugate frequencies. The optimal amplitudes of each mode, indicating its relative importance for the flow, were calculated as a best-fit of the data onto the DMD model \cite{jovanovic2014sparsity}.

\section{Results}

\begin{figure} 
    \centering
        \begin{subfigure}[b]{0.47\textwidth}
         \centering
         \includegraphics[width=\textwidth]{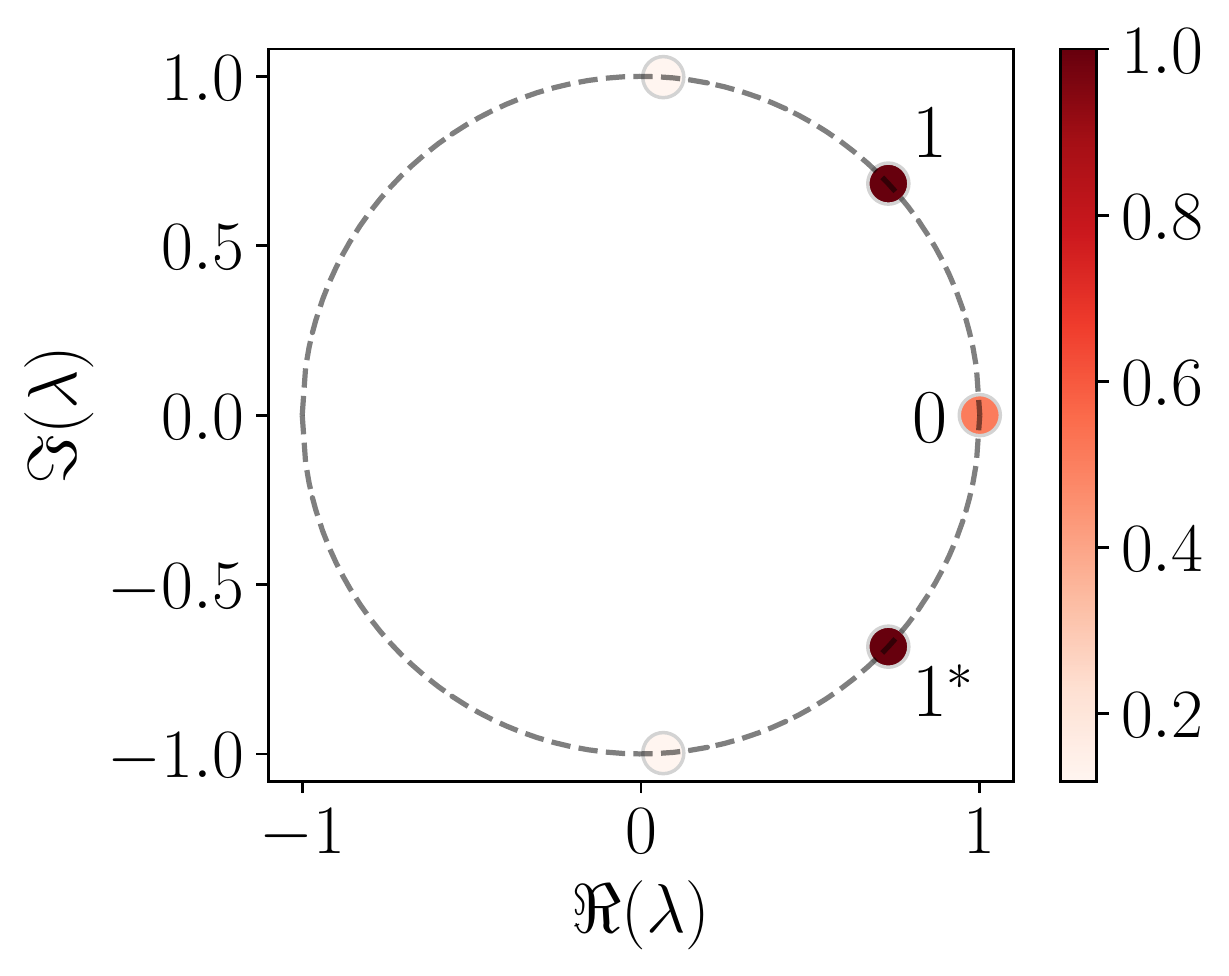}
         \caption{}
         \label{fig:lam_spec_sw}
     \end{subfigure}
     \hfill
      \begin{subfigure}[b]{0.47\textwidth}
         \centering
         \includegraphics[width=\textwidth]{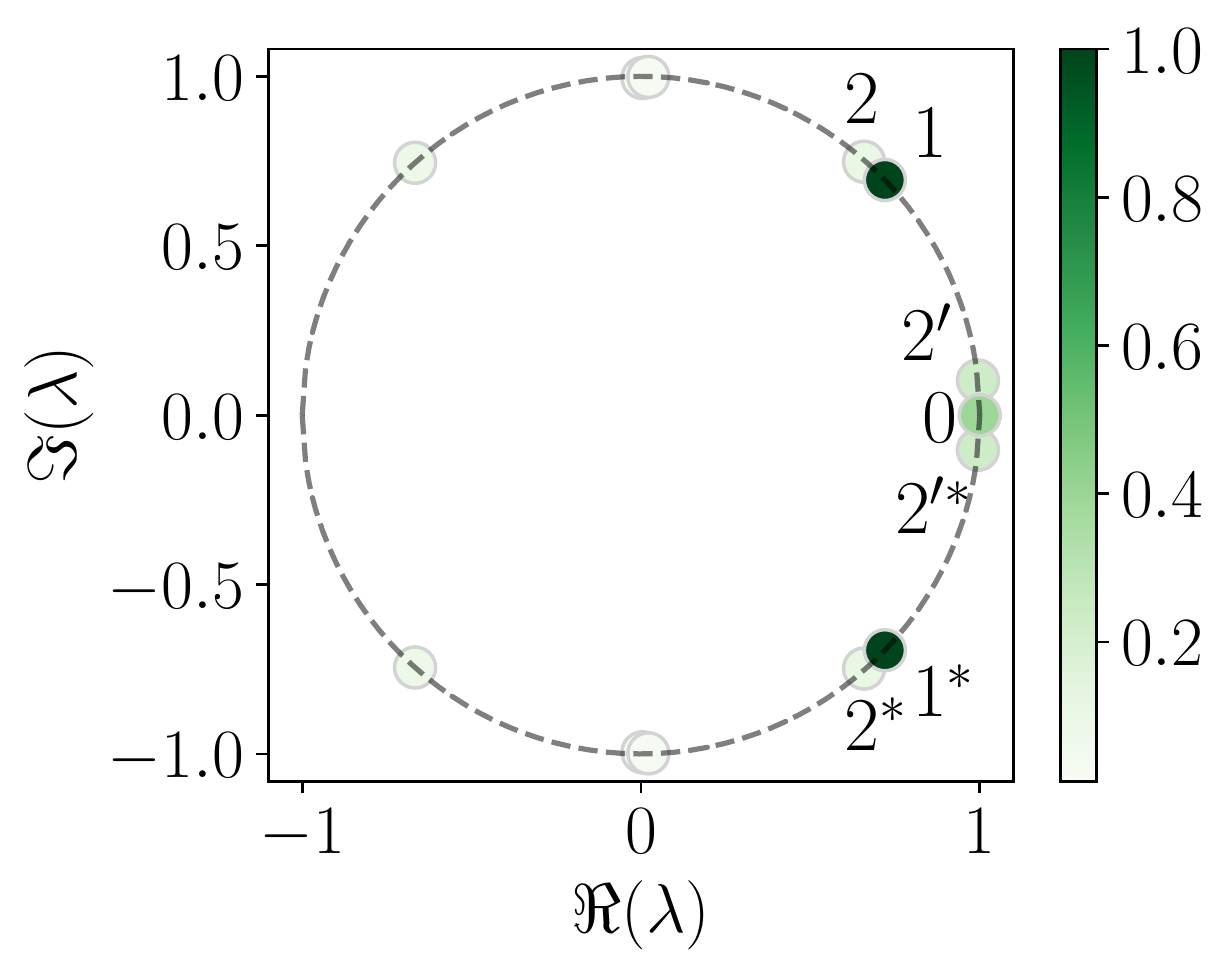}
         \caption{ }
         \label{fig:lam_spec_1f}
     \end{subfigure}
     \vfill
         \centering
        \begin{subfigure}[b]{0.47\textwidth}
         \centering
         \includegraphics[width=\textwidth]{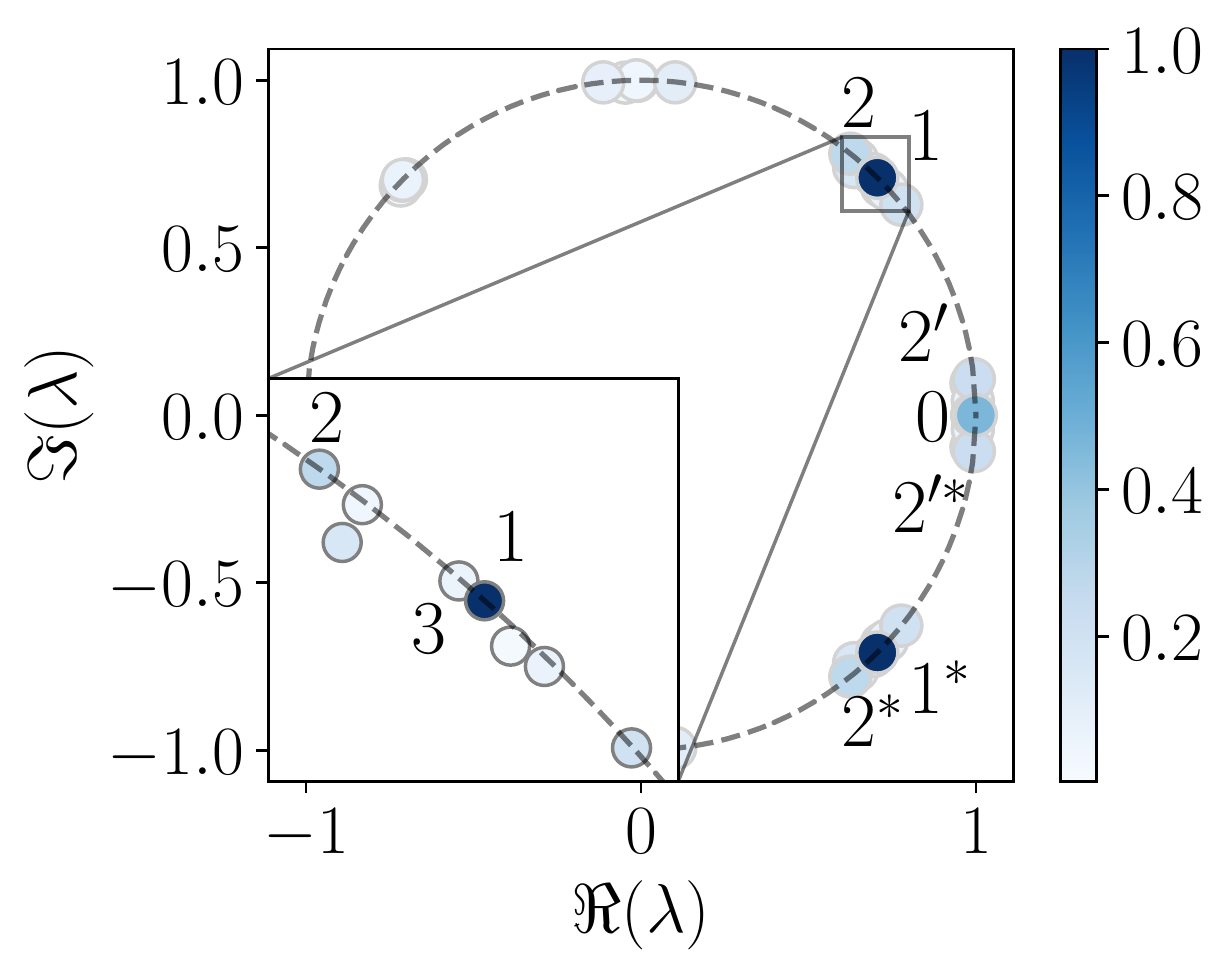}
         \caption{}
         \label{fig:lam_spec_2f}
     \end{subfigure}
     \hfill
      \begin{subfigure}[b]{0.47\textwidth}
         \centering
         \includegraphics[width=\textwidth]{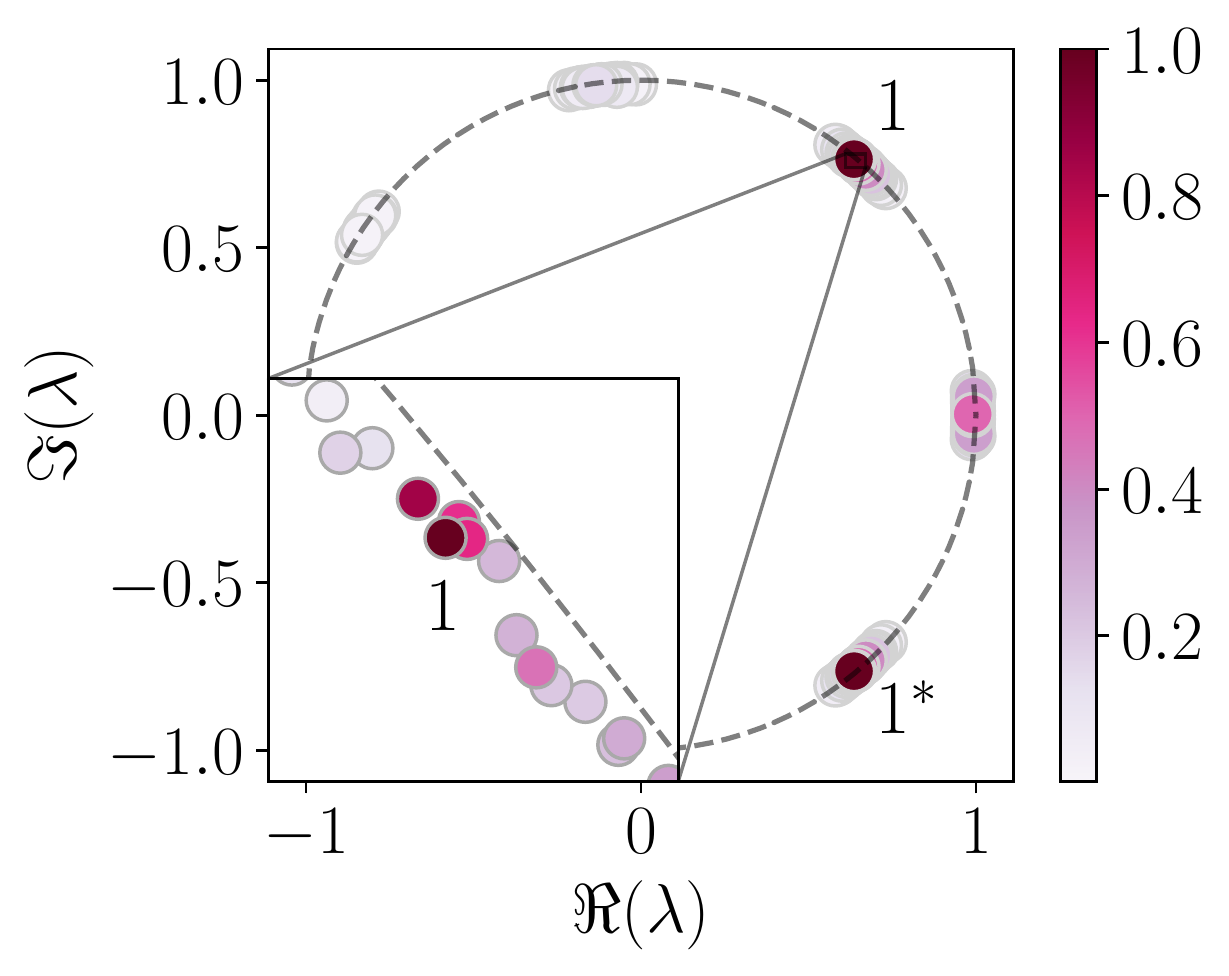}
         \caption{ }
         \label{fig:lam_spec_ch}
     \end{subfigure}
    \caption{The spectra of discrete DMD eigenvalues $\lambda$ of $u_z$, related to continuous-time eigenvalues as $\omega = \ln{\lambda}/\Delta t$, see~\eqref{eq:DMDcont}. (a) $Ha=149$, (b) $Ha=145$, (c) $Ha=140$, (d) $Ha=120$. The dashed circle  $|\lambda| =1$ corresponds to neutral stability. Color represents the optimal amplitudes of the modes, as defined by~\cite{jovanovic2014sparsity}. The modes identified as relevant for transition to chaos are numbered.}
    \label{fig:dmd_spec}
\end{figure}

In the following, we discuss our DMD results, corresponding to the dynamical regimes in table~\ref{tab:flow_regimes}. As the flow variables are related through \eqref{eq:divfree} and nonlinear terms in \eqref{eq:NSt}, \eqref{eq:Ind}, they have similar frequency content, so we first focus on the axial velocity $u_z$. Figure~\ref{fig:dmd_spec} presents its DMD spectra in the form of discrete eigenvalues~\eqref{eq:DMDdisc}. Figure~\ref{fig:mode_shapes} depicts the spatial structure of some of the modes of $u_z$ and $B_z$, and figure~\ref{fig:dmd_linstab_mean} compares DMD results to our linear stability analysis and DNS. We begin by discussing first the common features of the DMD spectra at different $Ha$, and then focus on the transition between the flow regimes.

\subsection{MRI and non-oscillatory modes}

\begin{figure}
     \centering
         \begin{subfigure}[b]{0.15\textwidth}
         \centering
         \includegraphics[width=\textwidth]{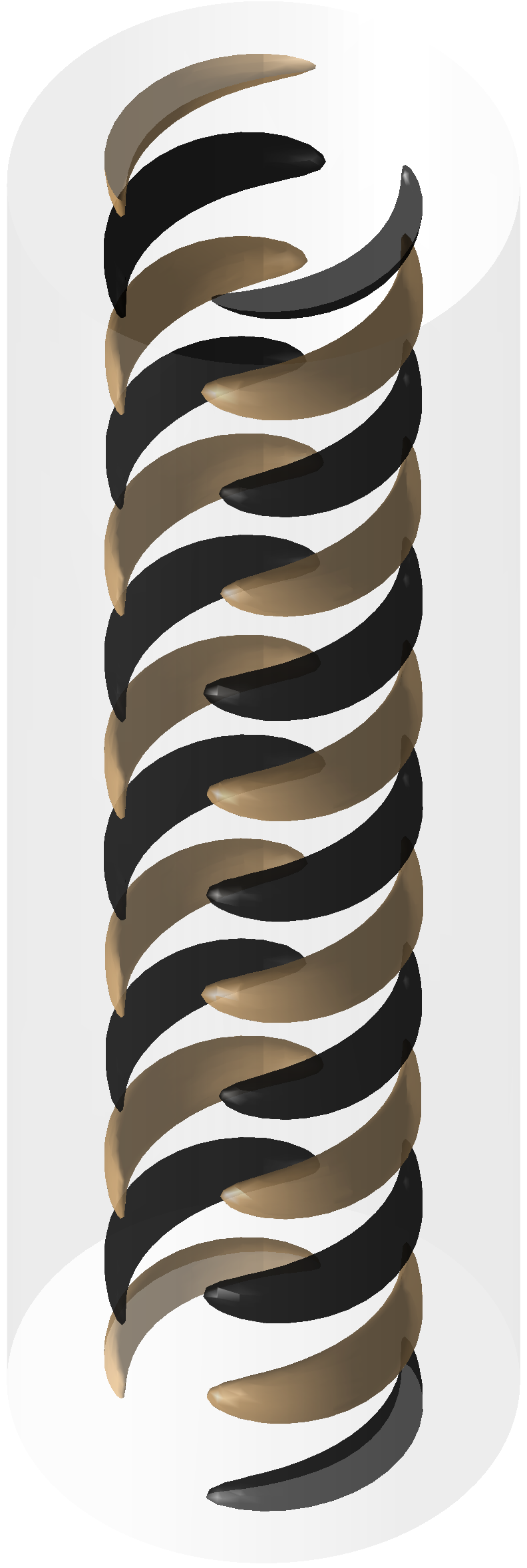}
         \caption{}
         \label{fig:vz_o1_mode}
     \end{subfigure}
    \hfill
         \begin{subfigure}[b]{0.15\textwidth}
         \centering
         \includegraphics[width=\textwidth]{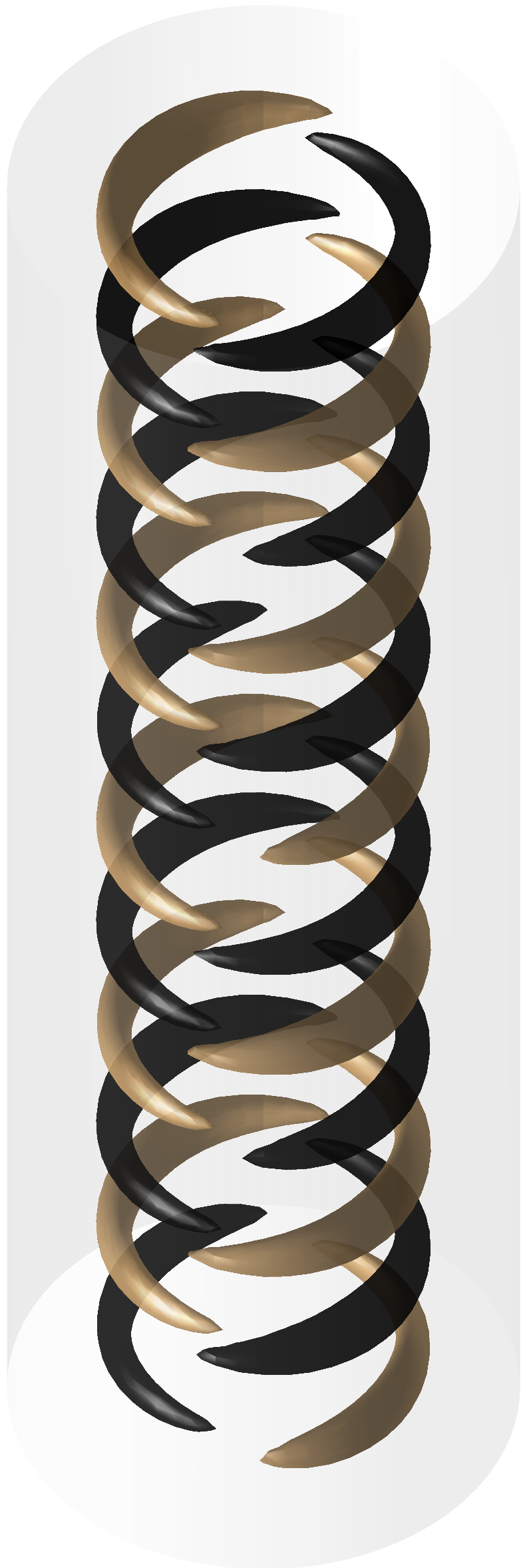}
         \caption{}
         \label{fig:vz_o2_mode}
     \end{subfigure}
     \hfill
         \begin{subfigure}[b]{0.15\textwidth}
         \centering
         \includegraphics[width=\textwidth]{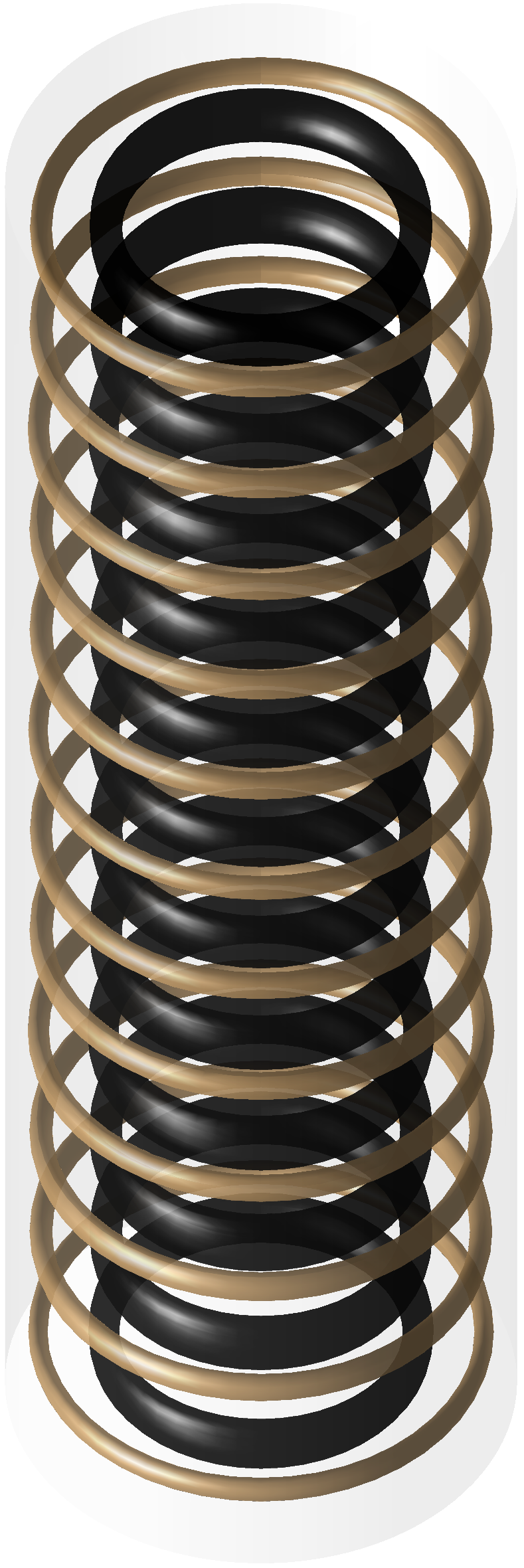}
         \caption{}
         \label{fig:vz_do1_mode}
     \end{subfigure}
          \hfill
         \begin{subfigure}[b]{0.135\textwidth}
         \centering
         \includegraphics[width=\textwidth]{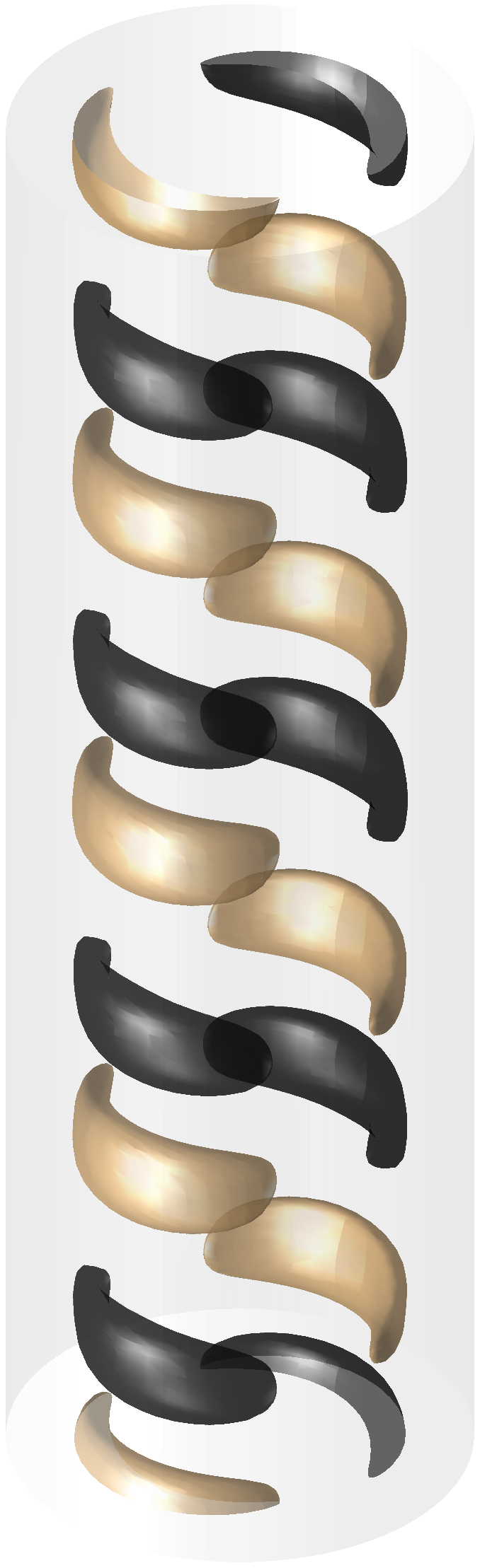}
         \caption{}
        \label{fig:bz_o2_mode}
     \end{subfigure}
         \hfill
         \begin{subfigure}[b]{0.135\textwidth}
         \centering
         \includegraphics[width=\textwidth]{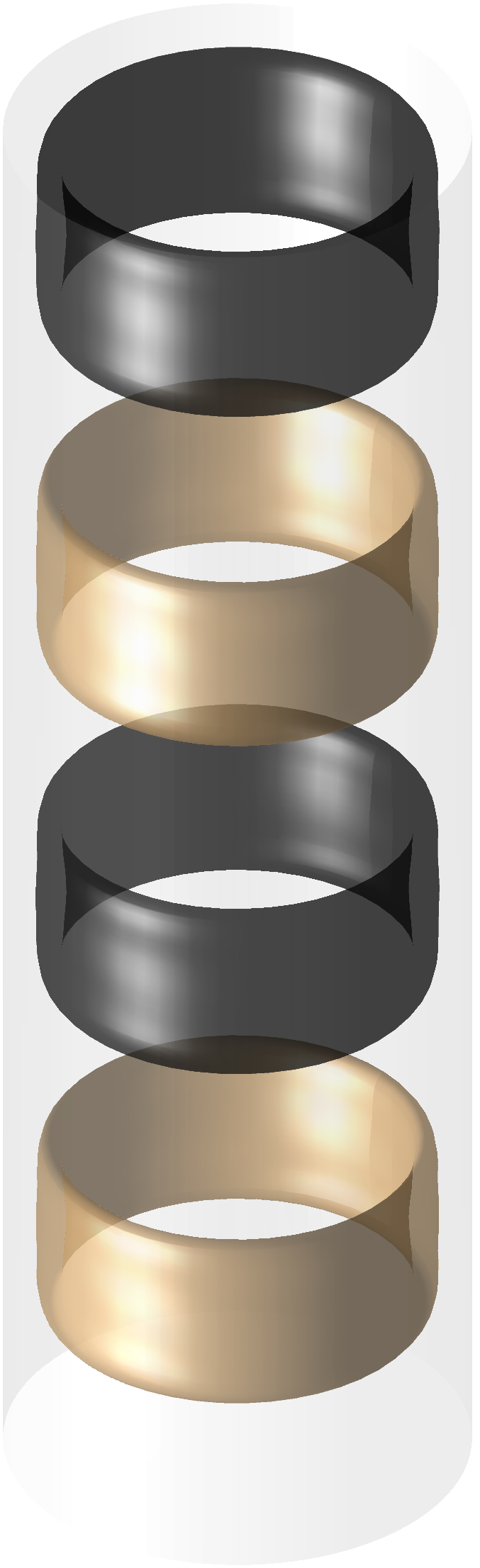}
         \caption{}
      \label{fig:bz_do2_mode}
     \end{subfigure}
        \caption{The shape of DMD modes identified as relevant for transition to chaos, $Ha=140$. Axial velocity $u_z$: (a) MRI wave, $\Im (\omega_1) \approx 0.3 \Omega_i$. (b) First splitting mode, $\Im(\omega_2) \approx 1.1 \Im(\omega_1)$.  (c) Mode resulting from  interaction of (a) and (b), with $\Im(\omega'_2) \approx \Im(\omega_2) - \Im(\omega_1)$. Axial magnetic field $b_z$: (d) Second  splitting mode, $\Im(\omega_3) = 1.025 \Im(\omega_1)$ (e) Interaction of (a) and (d), $\Im(\omega'_3) \approx \Im(\omega_3) - \Im(\omega_1)$.}
        \label{fig:mode_shapes}
\end{figure}

 In the DMD spectrum, the main oscillating component of the signal is represented by two modes with complex-conjugate  frequencies, denoted by $1$ and $1^*$ in figure~\ref{fig:dmd_spec}. This is the dominant mode for the explored values of $Ha$, and we denote its continuous-time analog as $\omega_1$. Its dominant wave number is $k = 3$, with six pairs of rolls along the domain length $L_z = 4\pi$  (figure~\ref{fig:vz_o1_mode}); it is non-axisymmetric with azimuthal wave number $m=1$, like the standing wave flow pattern in figure~\ref{fig:vz_iso_Ha149}. In figure~\ref{fig:omegai_vs_lin}, we compare the dominant DMD frequencies of $u_\phi$ \anna{and $B_\phi$}, $\Im(\omega_1)/\Omega_i \in (0.3,0.4)$, with the frequencies of the dominant MRI wave from the linear analysis. This frequency represents the azimuthal rotation of the MRI wave; both the linear and DMD frequencies decrease with $Ha$, \anna{with the latter slightly higher  up until $Ha\approx 120$. Then, as the emerging chaos becomes more pronounced,  the dominant DMD frequencies become smaller than the linear ones, especially for the modes of magnetic field.}  This frequency adjustment is expected, as the nonlinear saturation of the instability modulates its initial growth, and the chaotic flow includes several dynamical components with comparable frequency content (figure~\ref{fig:lam_spec_ch}). Despite the emerging chaotic motion, the agreement between the frequencies indicates that  the MRI-unstable modes remain active. The rest of the modes in figure~\ref{fig:dmd_spec} are clustered around $\omega_1$ and its higher harmonics, 
$\pm n \omega_1$, $n=2,3,...$. The latter, located on the left-hand side of the plots in figure~\ref{fig:dmd_spec}, arise due to self-interaction of the MRI modes in the nonlinear terms of~\eqref{eq:NSt},~\eqref{eq:Ind}, and have a finer spatial structure of $(k, m) = (6,2)$. They do not represent independent dynamics \anna{and are unlikely to play a role in the transition to chaos here, since it involves frequencies slower than the MRI ones}.

The non-oscillatory modes with $\Im(\omega_0)=0$, corresponding to a purely real discrete eigenvalue $\lambda=1$, are denoted by $0$ in figure~\ref{fig:dmd_spec}. \anna{They arise through the interaction of the two complex-conjugate MRI harmonics with $\pm\Im(\omega_1)$ and the temporal flow  mean}. This mode is axisymmetric in $\phi$ ($m=0$) with axial wave number $k = 6$. In the decomposition of $u_\phi$ and $B_\phi$ the $0$-mode corresponds to the axisymmetric mean flow, and has the largest amplitude;  its spatial structure in $z$ can be interpreted as a perturbation to the mean profile by the standing wave with $\omega_1$. By averaging this mode over $z$, we obtain the global mean profile of the DMD decomposition.
Figure~\ref{fig:mean_V_mode} compares this mean to the global spatiotemporal mean of the flow, and to the imposed laminar velocity profile~\eqref{eq:lamprof}. As MRI turbulence develops, the turbulent angular momentum transport modifies the imposed rotation profile; it become flatter in the bulk and develops large gradients at the walls.  This is not expected in a real astrophysical object, where turbulent fluctuations are thought to play a secondary role compared with gravity and mean rotation. The mean $\overline{B}_\phi$ flattens in a similar way and is also captured by DMD. 

\subsection{Periodic oscillation of torque and energies}
\begin{figure}
    \centering
    \begin{subfigure}[b]{0.47\textwidth}
         \centering
         \includegraphics[width=\textwidth]{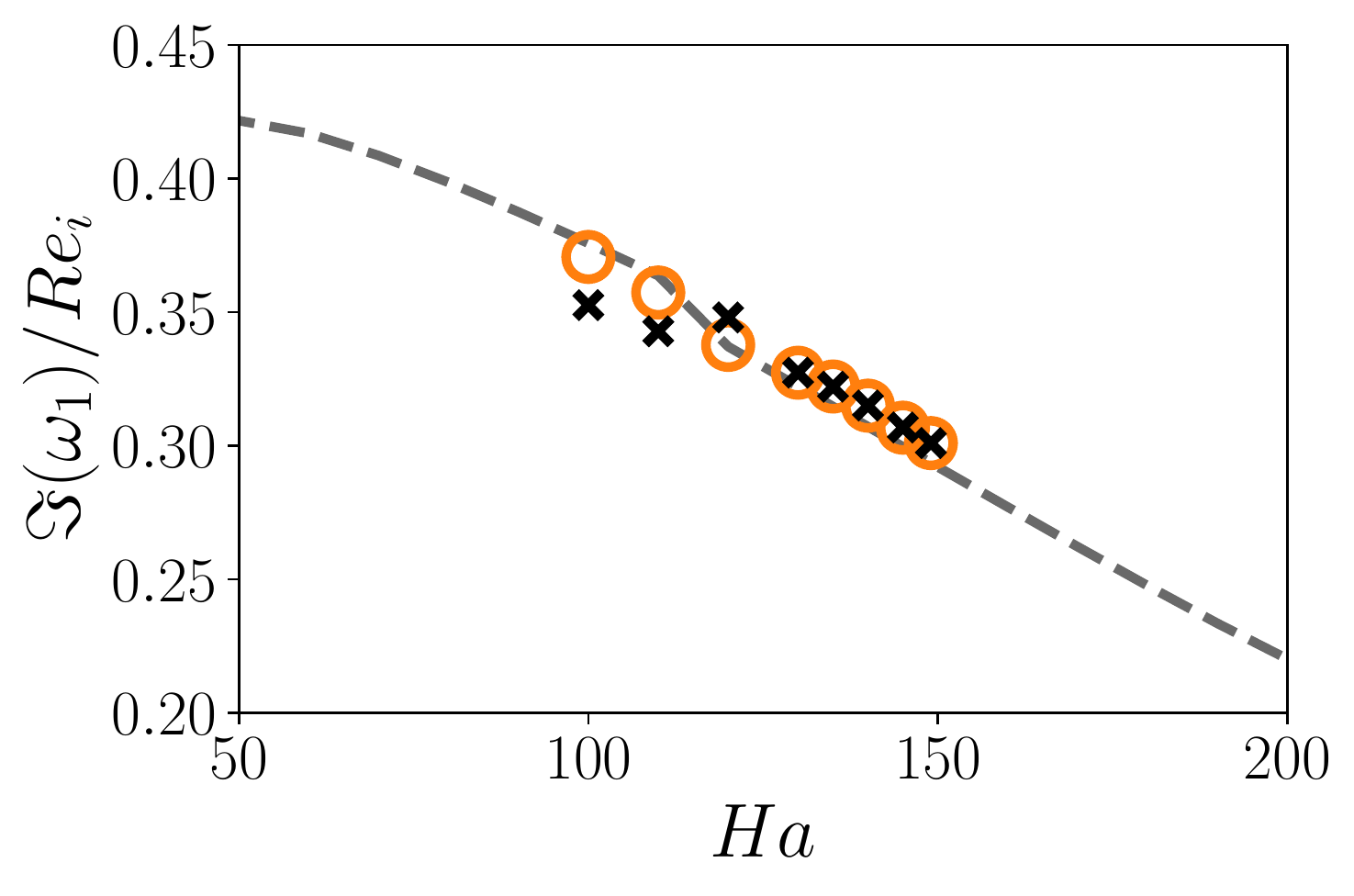}
         \caption{}
         \label{fig:omegai_vs_lin}
     \end{subfigure}
     \hfill
         \begin{subfigure}[b]{0.47\textwidth}
         \centering
         \includegraphics[width=\textwidth]{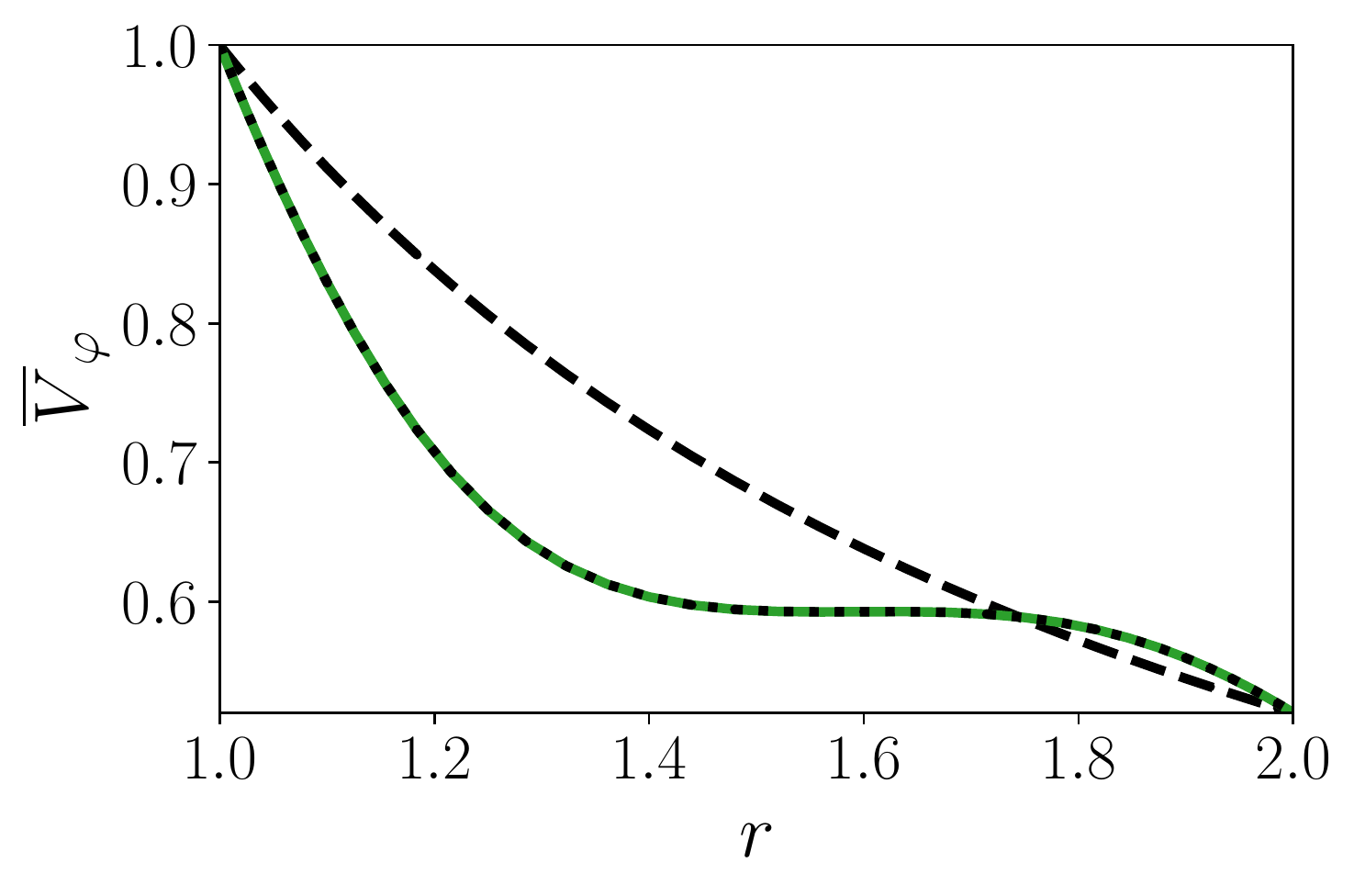}
         \caption{ }
         \label{fig:mean_V_mode}
     \end{subfigure}
     \vfill
         \begin{subfigure}[b]{0.45\textwidth}
         \centering
         \includegraphics[width=\textwidth]{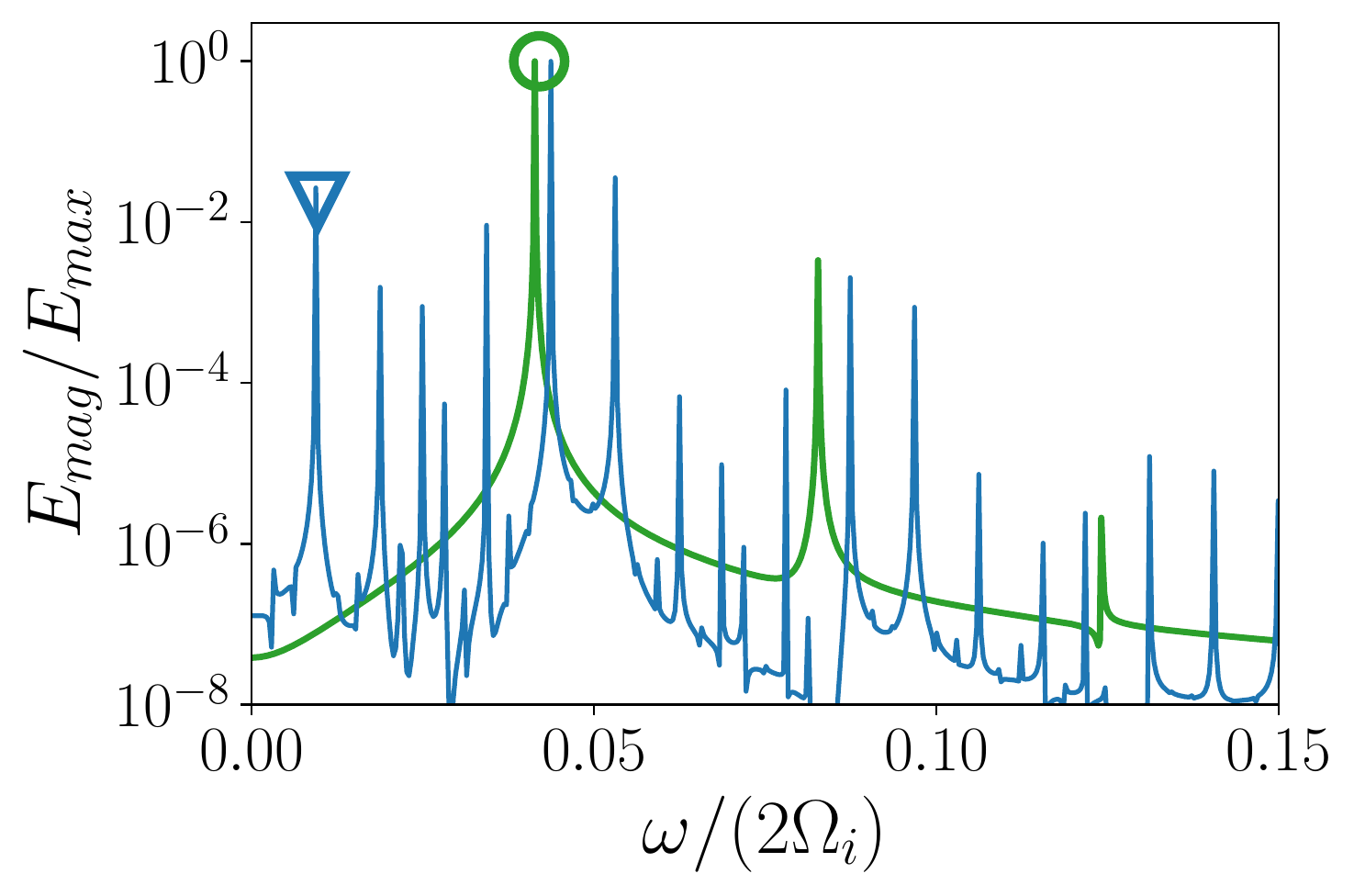}
         \caption{ }
         \label{fig:1tof2freq_fftG}
           \hfill
     \end{subfigure}
        \begin{subfigure}[b]{0.47\textwidth}
         \centering
         \includegraphics[width=\textwidth]{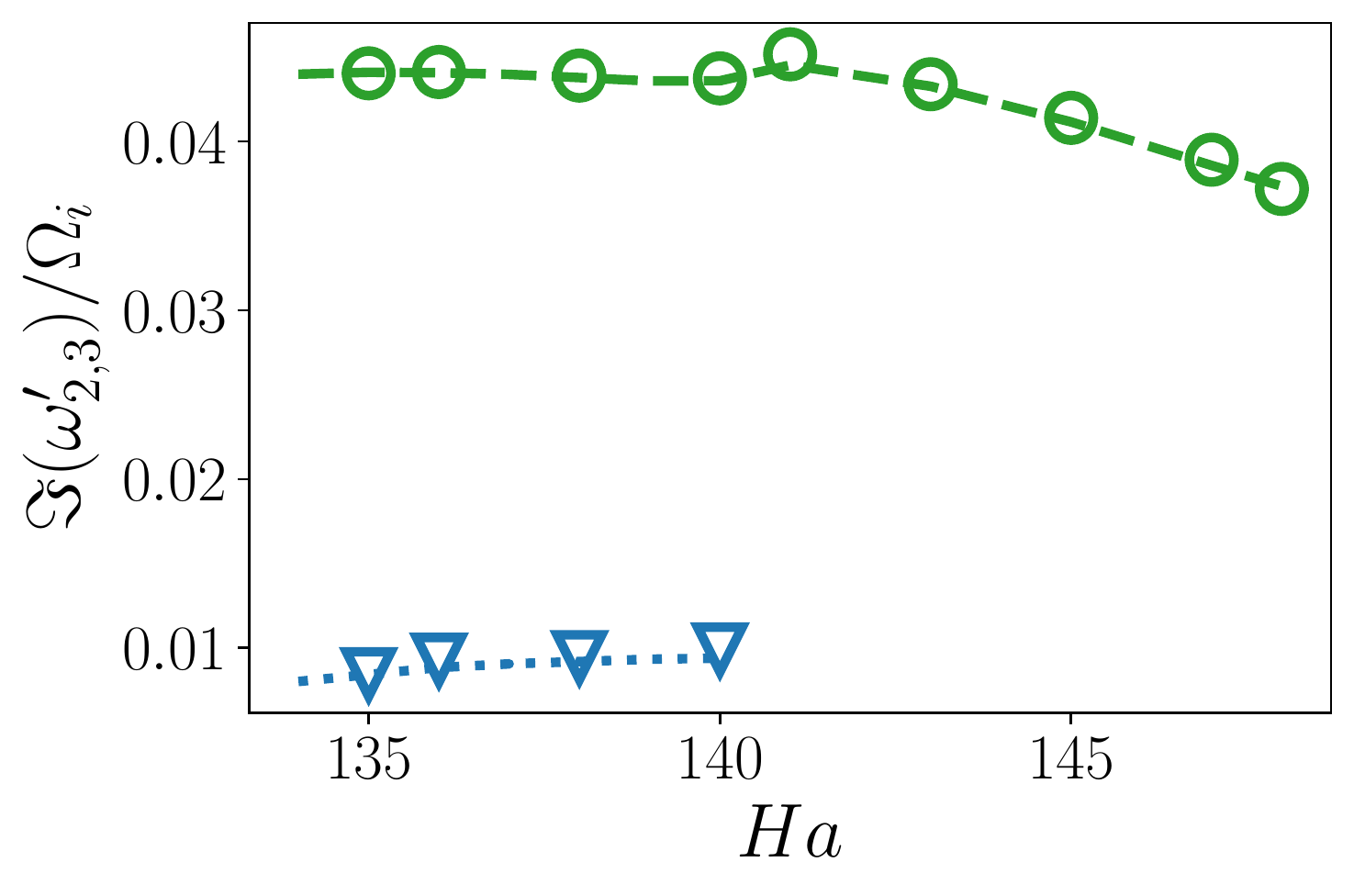}
         \caption{ }
         \label{fig:dmd2Gfreq}
     \end{subfigure}
  
    \caption{ (a) Comparison of DMD frequencies to the linear stability analysis. Dashed line, the linear MRI mode with largest growth; circles, the oscillating mode of $u_\varphi$, \anna{crosses, $B_\phi$}. (b) The flow mean, normalized with the velocity of the inner cylinder, at $Ha=145$ (color). Dotted line: the mean calculated from the dominant DMD mode of $u_\varphi$ with $\Im(\omega)=0$. Dashed line: laminar profile~\eqref{eq:lamprof}. (c) FFT of magnetic energy $E_{mag}$, with frequencies divided by a factor of $2$. Green, $Ha=145$ (as in figure~\ref{fig:flow_description}d); blue, $Ha=140$ (figure~\ref{fig:flow_description}e). \anna{The principal frequency of the energy} oscillation is denoted with a circle, and its modulating frequency with a triangle. (d) Comparison of the frequencies in panel (c) and the identified DMD frequencies.}
    \label{fig:dmd_linstab_mean}
\end{figure}

At $Ha=149$ only the MRI mode $\omega_1 \approx 0.3 \Omega_i$ and its harmonics are found in the flow (figure~\ref{fig:lam_spec_sw}). When the flow  enters the state with oscillating integral dynamics at $Ha=148$, two new modes simultaneously appear in the DMD decomposition (figure~\ref{fig:lam_spec_1f}). One of them, denoted by $2$ ($2^*$), has a frequency $\Im(\omega_2) \approx 1.1 \Im(\omega_1)$,  slightly faster than the frequency of the dominant mode. It also has the same periodicity in $z$ and $\phi$, $(k, m) = (3,1)$, as shown in figure~\ref{fig:vz_o2_mode}. In the following, we will refer to the appearance of modes with comparable frequency content as \textit{mode splitting}.  The second mode, $2'$ and $2'^*$, has a slower temporal evolution of $\Im (\omega'_2) \propto \Im(\omega_2) - \Im ( \omega_1)$, and is a result of interaction between the modes $1$ and $2$ through nonlinear terms. It has double periodicity in $z$, $k =6$, and is axisymmetric with $m=0$, as shown in figure~\ref{fig:vz_do1_mode}. This slow frequency, which has emerged in a secondary Hopf bifurcation, is the one forming the torus in figure~\ref{fig:vrtz_Ha147}, and is responsible for periodic oscillations of $G$ and flow energy $E$. In figure~\ref{fig:1tof2freq_fftG} we plot the Fast Fourier Transform (FFT)  of $E_{mag}$ of the flow for $Ha=145$, tracking this slower oscillation in the DNS. Note that both energies and torque are quadratic quantities~\eqref{eq:GJw}, so their frequencies are twice the frequencies detected in $\mathbf{u}$ and $\mathbf{B}$.  In the neighbouring panel~\ref{fig:dmd2Gfreq}, this frequency is compared with the slow frequency detected in the DMD decomposition of $B_z$.
In the respective region of $Ha$, both frequencies increase when $Ha$ decreases, with a good comparison between the two.

\subsection{Doubly-periodic oscillation and transition to chaos}
Now we consider the case of $Ha=140$ (figure~\ref{fig:lam_spec_2f}). As discussed before, with further decrease in $Ha$ wrinkles develop on the chaotic attractor until the flow becomes mildly chaotic (figure~\ref{fig:1tof2freq_poincare}). In this case, DMD decomposition becomes less robust but nevertheless it is possible to detect further mode splitting in figure~\ref{fig:lam_spec_2f}. Now both $\omega_1$ and $\omega_2$ are accompanied by neighbouring modes with comparable frequencies. The interaction of these modes leads to appearance of even slower modulations in the flow, with respective slow frequencies clustered around the mode $0$, behind  stronger signals from $\omega'_2$; this slow timescale may be the signature of the torus approaching a periodic orbit or fixed point in the flow.

In contrast to the purely periodic case with only one dominant frequency in $G$ and $E$, the FFT  reveals that here the signal is not perfectly doubly periodic and contains several frequency components. However, most of them can be identified as interactions between the previously detected frequency $\omega'_2 $ and the modulating one with $\Im(\omega'_3) \approx 0.01 \Omega_i$. In this regime, the former saturates at $\Im(\omega'_2)\approx 0.045 \Omega_i$, and the modulation becomes slower as  $Ha$ decreases (figure~\ref{fig:dmd2Gfreq}). We seek modes with a similar frequency component in the DMD decomposition of $b_z$, which has the lowest data rank compared to the rest of the flow for all $Ha$, and identify a second mode splitting of $\Im(\omega_3) \approx 1.025 \Im(\omega_1)$ (figure~\ref{fig:lam_spec_2f}). The new mode has a larger axial wave length with $(k,m)=(2,1)$ compared to the mode $1$ (figure~\ref{fig:bz_o2_mode}). It was absent in  more regular states of the flow and is not a result of harmonic self-interaction of the unstable modes, as its axial wave number is smaller. This mode is accompanied by a slow harmonic with $\Im(\omega'_3) \approx 0.01 \Omega_i$, which is a large-scale, axisymmetric structure of $k_z =1$, $m=0$ (see figure~\ref{fig:bz_o2_mode}), indicating triadic interaction among the modes $(\omega_1, \omega_3, \omega'_3)$. It is unclear whether mode $3$ or $3'$ is of the primary importance. In DMD of other flow variables, mode $3$ appears more consistently than  mode $3'$, and tends to have a higher optimal amplitude. In figure~\ref{fig:dmd2Gfreq}, we compare $\Im(\omega'_3)$ with the slow modulating frequency of $E_{mag}$, and observe that the two are in agreement.

\section{Discussion and outlook}
In this work, we have employed the data-driven analysis to track transition to chaos in Taylor--Couette flow subject to an azimuthal magnetic field. There, MRI arises as a standing wave through a supercritical Hopf bifurcation. In fluids with low conductivity (low $Pm$), a secondary subcritical Hopf bifurcation exists with an unstable edge state separating the periodic MRI and chaos \cite{guseva2015transition}. On the contrary, a fluid with high conductivity ($Pm=1$) shows more prolonged transition with diverse flow states \cite{guseva2015transition}. Here,
we focused on one region of this transition, $Ha \in (120,150)$, for fixed $Re=250$. With decreasing $Ha$, the friction on the cylinders changes from constant to oscillating, and then to a modulated signal, before becoming chaotic. This transition seemingly follows a well-known Ruelle-Takens scenario, with a cascade of two Hopf bifurcations, the first at the onset of MRI, and the second when periodic oscillations of $G$ and $E$ develop. In the phase space, it involves a periodic orbit, a torus, and then a breakdown of the torus through folding and wrinkling of the attractor (figure~\ref{fig:phase_space}). On the other hand, the alignment of the attractor remains relatively unchanged, despite developing chaotic dynamics.

We employed Dynamic Mode Decomposition to identify the changes in the flow responsible for the temporal behaviour of its friction and energy. The first transition from the MRI standing wave (mode $1$) to periodic oscillations in $G$ happens in a process of mode splitting, i.e. mode $2$ similar to the MRI mode with a slightly different frequency appears in the domain.  Intuitively, the appearance of the oscillation in $G$ and $E$ in this case can be understood as a symmetry breaking in the system. A flow with two  dynamical components,  rotating in $\phi$ at different frequencies, is no longer invariant in $\phi$ in terms of the integral quantities. The friction on the cylinders at any time  depends on the particular alignment of the two dynamical flow structures, periodically returning to their initial configuration.

The next flow state, where $G$ is modulated by a slower frequency, is mildly chaotic in $r$- and $\phi$-directions of the flow, however, it remains relatively ordered in the $z$ direction. The DMD decomposition of $b_z$ shows the appearance of new modes $3$ and $3'$ with larger axial wave lengths; their footprint is visible in the rest of the flow variables. They create a frequency content comparable to the modulating frequency of the torque. The frequency $\Im(\omega'_3)$ is related to positive and negative regions of the field in figure~\ref{fig:bz_do2_mode} interchanging their location along $z$. As the flow becomes more complex in this case, with the modes $1$, $2$, $2'$ also influencing dynamics (figure~\ref{fig:lam_spec_2f}), DMD detects the slow modulation as a set of modes of similar frequency content and spatial shape, slightly different for different flow variables. Although we attribute the modulation in $G$, $E_{kin}$ and $E_{mag}$ to the presence of modes $3$ and $3'$, the modulation of $G$ and $E$ is likely a cumulative effect of all these harmonics. On the other side, the $\omega'_3$ modal component of magnetic field with $k = 1$ and $k=2$ is clearly present in the dynamics of magnetic field in figure~\ref{fig:bz_iso_Ha100_k4} and also in its space-time plots (not shown here).  Thus, DMD was able to identify the flow components relevant for transition to chaotic dynamics.

As the transition to chaos proceeds further, DMD represents the flow as a set of splitting frequencies clustered about the originally dominant MRI modes (figure~\ref{fig:lam_spec_ch}). There are still only a few modes with large amplitudes, highlighting the low-dimensionality of this chaotic attractor. The DMD modes of chaotic flow lie inside the unit circle and appear dampened. However, their instantaneous temporal coefficients (not shown here) have chaotic rather than decaying dynamics, indicating that the linearity assumption of~\eqref{eq:DMDcont} is no longer valid and nonlinear dependencies between the modes emerge. This is not a concerning issue here, since we used DMD not for reduced-order modelling of the system, but as a diagnostic tool for its dynamical behaviour. The future work will include relating the temporal evolution of the modes into a nonlinear model, together with improving robustness of the presented DMD method by taking into account flow symmetries  \cite{baddoo2021physics,marensi2021symmetry}, or harnessing statistical properties of the flow \cite{sashidhar2022bagging}. \anna{Such model could provide a quantitative description of nonlinear interactions accompanying the transition to MRI turbulence.} \\

\textit{This project has received funding from the European Union’s Horizon 2020 research and innovation programme under the Marie Skłodowska-Curie grant agreement No 890847.}

\bibliographystyle{ieeetr}
\bibliography{tcfbib} 

\end{document}